\documentclass[]{aa} 
\usepackage{natbib}
\usepackage{graphicx}
\usepackage{txfonts}
\def\A2163B{$\mathrm{A\,2163B}$} 
\def\A2163{$\mathrm{A\,2163}$} 
\def\lre{$\mathrm{\log R_e}$}
\def\lRe{$\mathrm{\log R_e}$}
\def\re{$\mathrm{R_e}$}

\def\mie{$\mathrm{< \! \mu \! >_e}$} 
\def\mtot{$\mathrm{m_T}$}
\def\ns{$\mathrm{n}$}
\def\lns{$\mathrm{\log n}$}
\begin{document}
\titlerunning{The cluster of galaxies $\mathrm{ A\,2163B }$.}

\title{Probing galaxy evolution through the internal colour gradients,
     the  Kormendy  relations and  the  Photometric  Plane of  cluster
     galaxies  at  $\rm  z  \sim  0.2$.\thanks{Based  on  observations
     collected  at  European   Southern  Observatory  (ESO  65.O-0251,
     69.D-0653).  Tab.~2 is fully  available in electronic form at the
     CDS via  anonymous ftp  to cdsarc.u-strasbg.fr.  }   } \author{F.
     La   Barbera  \and   P.    Merluzzi  \and   G.   Busarello   \and
     M. Massarotti \and A.  Mercurio }

\offprints{F. La Barbera}

 \institute{ I.N.A.F., Istituto Nazionale di Astrofisica Osservatorio
 Astronomico di Capodimonte, Via Moiariello 16, I-80131 Napoli \\
 email: labarber@na.astro.it  
}

   \date{Received ; accepted }

   \abstract{We  present  a   detailed  analysis  of  the  photometric
   properties of  galaxies in the  cluster \A2163B at redshift  $\rm z
   \sim  0.2$. R-, I-  and K-band  structural parameters,  (half light
   radius  \re, mean  surface brightness  \mie  \, within  \re \,  and
   Sersic index \ns) are derived for $\rm N \sim 60$ galaxies, and are
   used to study their internal colour gradients.  For the first time,
   we use the slopes of optical-NIR Kormendy relations to study colour
   gradients  as  a  function  of  galaxy  size,  and  we  derive  the
   Photometric  Plane  at $\rm  z\sim  0.2$  in  the K  band.   Colour
   gradients are  negligible at optical wavelengths,  and are negative
   in the optical-NIR,  amounting on average to $\rm  -0.48 \pm 0.06$.
   This  result is  in  agreement with  our  previous measurements  of
   colour gradients at intermediate redshifts, and imply a metallicity
   gradient in  galaxies of $\rm \sim0.2~dex$ per  radial decade.  The
   analysis of the Kormendy relation suggests that its slope increases
   from the optical to the  NIR, implying that colour gradients do not
   vary or even do become less steep in more massive galaxies.  Such a
   result  is  not simply  accomodated  within  a monolithic  collapse
   scenario,  while it can  be well  understood within  a hierarchical
   merging framework.   Finally, we  derive the first  NIR Photometric
   Plane at $\rm z \sim  0.2$, accounting for both the correlations on
   the  measurement  uncertainties  and  the selection  effects.   The
   Photometric Plane  at $\rm z \sim  0.2$ is consistent  with that at
   $\rm  z \sim 0$,  with an  intrinsic scatter  significantly smaller
   than the Kormendy relation but larger than the Fundamental Plane.

 }

   \maketitle
%

\section{Introduction}

Studies of  galaxy populations  in clusters at  intermediate redshifts
have been proved  to be an effective tool  to constrain the mechanisms
underlying their formation and evolution.  A pure photometric approach
involves the analysis of  galaxy Luminosity Functions (LFs), of galaxy
colours, mainly  via the  colour magnitude (CM)  diagrams, and  of the
internal  light  distribution  of  galaxies,  i.e.   their  structural
parameters.

The galaxy  LF and  its dependence on  the waveband  carry information
both  on  the stellar  processes  occurring  in  galaxies and  on  the
cosmological  processes which  determine their  mass  function.  Since
\citet{PrS74}, the  shape of the LF has  been seen as a  basic test of
theories of  structure formation  and evolution.  The  luminosities of
early-type galaxies  are also  tightly correlated with  their colours,
through the colour magnitude  (CM) relation \citep{ViS77, BLE92}.  The
slope of this relation shows little evolution with redshift up to $\rm
z \sim 1$,  while its zeropoint changes according  to what is expected
for   an    old   passively   evolving    stellar   population   (e.g.
\citealt{SED98}, \citealt[hereafter KAB98]{KAB98}).  As shown by KAB98
(see  also \citealt{MLM03}),  the evolution  of the  CM  sequence with
redshift  sets strong  constraints  on the  origin  of this  relation,
resolving the well known age--metallicity degeneracy \citep{WTF96}.

The structural properties of galaxies at different redshifts provide a
wealth of  information by which galaxy evolution  can be investigated,
both through the study of the correlations and through the analysis of
the waveband dependence  of such quantities.  The NIR  light, in fact,
is  less  sensitive to  more  recent  star  formation, to  metallicity
effects  (though line  blanketing) and  to dust  absorption, following
more closely the luminous matter distribution.

The multi-waveband  analysis gives information on  the internal colour
gradients (CGRs) of galaxies, and therefore on the radial gradients of
the  properties  of their  stellar  populations  (SPs),  such as  age,
metallicity  and  dust  content.   Nearby  galaxies  have  on  average
negative  CGRs,   their  SPs  becoming  bluer   toward  the  periphery
\citep{PDI90, PVJ90}.   Due to the  low evolution of  colour gradients
with redshift,  metallicity seems to be  the primary driver  of the SP
gradients   in   spheroidal   galaxies,   as  demonstrated   both   by
optical--optical  studies   (e.g.   \citealt{SMG00,  TaO00})   and  by
optical--NIR CG measurements \citep[hereafter LBM02 and LBM03a]{LBM02,
LBM03a}.   Age gradients,  however, are  not  fully ruled  out by  the
present  data (\citealt{SMG00},  LBM03a),  while the  effects of  dust
absorption  still remain  substantially unresolved.   The  presence of
metallicity  gradients  can  be   well  explained  in  the  monolithic
formation scenario  of early-type  galaxies \citep{LAR74}, due  to the
later  beginning of  a  galactic  wind in  the  inner galaxy  regions.
However, it  can also be accommodated within  the hierarchical merging
framework, since metallicity gradients are settled in disk galaxies at
high redshifts, and more massive  early-types form on average from the
merging of larger disks \citep{KAU96}.

Structural parameters, such as the effective (half--light) radius \re,
the mean surface brightness \mie \, within \re \, and the Sersic index
\ns,  are  correlated by  various  relations,  whose physical  origins
reside  both in  the properties  of  the SPs  of galaxies  and in  the
dynamical structure  of these  systems.  For the  early-type galaxies,
one of these correlations is that between \lre \, and \mie, also known
as  the Kormendy  relation (KR).   As  shown by  Capaccioli, Caon  and
D'Onofrio~(1992), $\mathrm{early-type \,  galaxies + bulges}$ form two
distinct families  in the plane  of effective parameters: that  of the
bright early-types,  following the KR, and  another `ordinary' family,
whose  properties  are  more  disperse and  heterogeneous.   Recently,
\citet{GrG03} showed that a real  dichotomy does not exist between the
two families,  but their observed  properties are well explained  by a
systematic change  in the profile  shape with galaxy  luminosity.  The
evolution of the zeropoint of  the KR has been largely investigated at
optical wavebands to constrain the major formation epoch of galaxy SPs
and  to  perform  the  Tolman  test  for  the  cosmological  expansion
\citep{SaP91, PDdC96, BAS98, ZSB99, SaL01, LuS01}.  On the other hand,
the slope  of the KR is an  interesting tool to gain  insight into the
properties  of   the  galaxy  SPs   as  a  function  of   galaxy  size
(\citealt{ZSB99},  \citealt{LBM03b}  hereafter  LBM03b),  which  is  a
crucial  prediction of  hierarchical merging  scenarios.   As recently
shown  by  \citet[hereafter  GRA02]{GRA02}, early-type  galaxies  also
follow a  three-dimensional relation between  \lre, \mie \,  and \lns,
which  is similar to  the spectroscopic  Fundamental Plane  (FP), once
velocity   dispersions   are  replaced   by   Sersic  indices.    This
`photometric plane' (hereafter PHP) has an observed scatter comparable
to  that   of  the  Fundamental  Plane  (FP)   relation,  and  carries
interesting  information on  the physical  properties of  galaxies and
their origin.

In the framework of a  project aimed at investigating the optical--NIR
photometric  properties  of  galaxies  in  the fields  of  ROSAT  PSPC
extended sources,  we have obtained multi-waveband data  (BVRIK) for a
field of $5'\times5'$ centered at $\rm RA= 16h \, 15m \, 48s$ and $\rm
DEC=-06^\circ  \, 02'  \, 10''  $, at  about $6.5'$  North  ($\rm \sim
1.3~Mpc$ at  $\rm z\sim0.2$ with $\mathrm{  H_0= 70~Km~s^{-1} Mpc^{-1}
}$,  $\Omega_m=0.3$ and  $\Omega_\Lambda=0.7$)\footnote{We  adopt this
cosmology throughout the paper.  With  these parameters the age of the
universe  is  $\mathrm{\sim13.5~Gyr}$,  and  the  redshift  $\rm  0.2$
corresponds to  a lookback time  of $\mathrm{\sim 2.5~Gyr}$.}   of the
Abell  cluster \A2163  at redshift  $\rm z=0.201$  \citep{AEB94}.  The
coordinates  of this  field correspond  to the  center of  a secondary
extended  emission in the  X-ray map  of \A2163,  known as  \A2163B \,
\citep[hereafter EAB95]{EAB95},  with an  X-ray temperature $\rm  KT >
6~keV$.  The  cluster \A2163  \, ($\rm  RA= 16h \,  15m \,  56s$, $\rm
DEC=-06^\circ \, 08' \, 55'' $)\footnote{The coordinates refer to $\rm
J2000$ and define the peak of  the X-ray emission (EAB95).}  is a very
rich  and complex structure  of galaxies,  which has  been extensively
studied  for its  exceptionally hot  X-ray temperature,  $\rm  KT \sim
12-15 keV$ (see \citealt{MMI96},  and references therein), and for its
huge   radio  halo,   one  of   the   most  powerful   known  so   far
\citep[hereafter  FFG01]{FFG01}.   Our data  show  the  presence of  a
significant excess of galaxies at  redshift $\sim 0.2$ in the field of
\A2163B,  with a  NIR spatial  distribution typical  for a  cluster of
galaxies, with a main central overdensity and a secondary structure at
$\rm \sim0.4~Mpc$ in the North-East direction.

In  the  present  work,   we  study  the  photometric  and  structural
properties  of the  galaxy population  of \A2163B,  by  discussing the
constraints  implied  by our  results  on  different galaxy  evolution
scenarios.

The  layout of  the paper  is as  follows. The  data are  presented in
Sec.~\ref{THEDATA},  while  the  luminosity  density map,  the  colour
magnitude relations  and the  K-band LF of  \A2163B \, are  studied in
Sec.~\ref{THECLUSTER}.   Cluster  members  are  then selected  by  the
photometric redshift technique, as described in Sec.~\ref{ZPSEC}.  The
derivation  of structural  parameters is  outlined in  Sec.~5.  Sec.~6
deals  with  the comparison  of  structural  parameters  and with  the
analysis of the internal  colour gradients of galaxies.  Secs.~7 and~8
deal  with the  correlations  between the  structural parameters.   In
Sec.~7,  we attempt  for  the  first time  to  perform an  optical/NIR
comparative study  of the \lre--\mie  \, relation, while in  Sec.~8 we
use the K-band data to analyze the PHP for the population of spheroids
in \A2163B.  Discussion and  conclusions follow in Sec.~9.  Details on
data  reduction are  given in  Appendix~A, while  the estimate  of the
cluster  redshift from the  optical--NIR CM  relations is  detailed in
Appendix~B.  The  catalogue, including all  the photometric properties
of galaxies in  the field of \A2163B, that  is BVRIK total magnitudes,
colours,  photometric  redshifts  and  RIK structural  parameters,  is
described in Section~2.3.


\section{The data}
\label{THEDATA}
\subsection{Observations}
The data relevant for the present  study were collected at the ESO New
Technology  Telescope and  include  BVRIK photometry  centered on  the
field  of galaxies  \A2163B.  The  data  were obtained  on August  and
September  2000  (hereafter  run  I),  under  partly  non  photometric
conditions, and on April 2002 (run II), under photometric conditions.

During three nights of run I, we obtained K-band imaging with the SOFI
instrument (pixel scale  $\rm 0.288''/pxl$) for a field  of $5' \times
5'$.   The sky  conditions were  photometric  for two  nights and  non
photometric  for the  third  night.   A total  of  120 exposures  were
collected, with a DIT of $\mathrm{  6~s}$, $\mathrm{ NDIT = 10 }$, and
a  dithering  box  of  $\mathrm{18''}$.  Due to  the  non  photometric
conditions,  only 85  exposures were  retained, resulting  in  a total
integration   time   of   $\mathrm{5100~s}$.   Standard   stars   from
\citet{PMK98} were observed during  each night, each at five different
positions  on  the  chip.    In  order  to  perform  the  illumination
correction (see Sec.~\ref{KDATA}), we also observed a standard star at
different positions on  a $5 \times 5$ grid across  the frame.  In the
same period, we  also obtained V-, R-, and  I-band photometry with the
EMMI instrument for  a region of $8.6' \times  8.6'$ (pixel scale $\rm
0.267''/pxl$), under  non photometric conditions.  On  April 2002, the
field \A2163B  was re-observed in B, V,  R and I band  with EMMI under
photometric conditions.  The  relevant information on the observations
are summarized  in Tab.~\ref{OBS}. The total exposure  times amount to
$600$, $3000$,  $1200$ and $\rm 1200~s$ for  the B, V, R  and I bands,
respectively.
\begin{table}
  \caption[]{Observations. The notes refer to the sky conditions: $\rm
  NP$ and  $\rm P$ denote non-photometric  and photometric conditions,
  respectively.}
  \label{OBS}
  $$ 
  \begin{array}{lcccccc}
    \hline
    \noalign{\smallskip}
     \rm Run    &  \rm Date  & \rm Note & \rm Waveband & \rm N_{exp} & \rm T_{exp}  & \rm FWHM \\
         &    &  &  &  & (s) & ('') \\
    \noalign{\smallskip}
    \hline
    \noalign{\smallskip} 
       I & 25/08/2000       & \rm NP & \rm V & 3  & 900 & 1.3 \\
       I & 24/08/2000       & \rm NP & \rm R & 2  & 450 & 1.0 \\
       I & 26/08/2000       & \rm NP & \rm I & 2  & 450 & 1.0 \\
       I & 08,10/09/2000    & \rm P  & \rm K & 70 &  60 & 0.6-1.0 \\
       I & 09/09/2000       & \rm NP & \rm K & 50 &  60 & 1.0 \\
      II & 09/04/2002       & \rm P  & \rm B & 2  & 300 & 1.5 \\
      II & 09/04/2002       & \rm P  & \rm V & 1  & 300 & 1.2 \\
      II & 09/04/2002       & \rm P  & \rm R & 1  & 300 & 0.9 \\
      II & 09/04/2002       & \rm P  & \rm I & 1  & 300 & 1.0 \\
    \noalign{\smallskip}
    \hline
  \end{array} 
  $$ 
\end{table}
\subsection{Analysis}
\label{APSEC}
The  data were  reduced by  using  FORTRAN routines  developed by  the
authors and IRAF packages\footnote{IRAF is distributed by the National
Optical Astronomy Observatories, which are operated by the Association
of  Universities for  Research in  Astronomy, Inc.,  under cooperative
agreement with the National Science Foundation.  }.  Details are given
in Appendix~\ref{BVRIDATA} and~\ref{KDATA} for the optical and the NIR
images, respectively.   Photometry was  derived by using  the software
S-Extractor  \citep{BeA96}.  For the  optical images,  S-Extractor was
run in the so-called double-image mode, using the R-band image for the
detection of the  sources, and the other images  only to measure their
fluxes.   Particular  care  was  taken  so that  closed  objects  were
deblended in the  same way both in the optical and  in the NIR images.
The BVRI and the K-band catalogues were cleaned by spurious detections
and cross-correlated.  For each  object, we computed colour indices by
using  magnitudes  within  a   fixed  aperture  of  diameter  $3.5''$,
corresponding  to $\rm \sim  12~kpc$ at  redshift $z=0.2$.   Since the
images have  different seeing, the aperture  magnitudes were corrected
to  the same  FWHM of  $1.2''$,  following the  procedure detailed  in
\citet{LMI03c}.   The aperture  corrections amount  to  $-0.04$, $0.$,
$0.02$, $0.02$,  and $\rm 0.04~mag$  for the B,  V, R, I and  K bands,
respectively.   Total magnitudes  were obtained  with  the S-Extractor
mag-auto parameter, which provides  magnitudes in an adaptive aperture
of diameter  $\mathrm{\alpha \cdot r_K}$, where  $\mathrm{r_K}$ is the
\citet{KRO80} radius,  and $\alpha=2.5$, for which  the Kron magnitude
encloses $94  \%$ of the object  source.  Total magnitudes\footnote{In
the following, total magnitudes will be indicated by a pedex $\rm T$.}
were  then  obtained  by  subtracting  $\rm 0.06~mag$  from  the  Kron
magnitudes. The  uncertainties on  galaxy magnitudes were  obtained by
adding in quadrature  the errors on the photometric  zeropoints to the
uncertainties  estimated by  S-Extractor.  Galaxy/star  separation was
performed on the  basis of the stellar index  $\rm SG$ of S-Extractor,
defining as stars the objects with $\rm SG \ge 0.97$. The completeness
of the K-band  photometry, which is used in  Sec.~\ref{LFKSEC} for the
LF  analysis, was derived  by following  the method  of \citet{GMA99}.
This method  consists in the  determination of the magnitude  at which
objects start to be lost  since they are below the brightness treshold
in  the  detection cell.   To  estimate  the  completeness limit,  the
magnitudes in  the detection  cell were computed  and compared  to the
total magnitudes.  This approach allows to define a conservative value
for  the  completeness  magnitude,  corresponding  to  a  completeness
percentage of $\sim 100 \%$.  The completeness magnitude in the K band
is $\rm  K_T=19$.  Down  to this limit,  the BVRIK  catalogue includes
$\rm N=317$  galaxies.  The completeness  of the BVRIK  catalogue with
respect  to the  K-band  catalogue is  $\sim98\%$  at $\rm  K_T=17.5$,
falling  to $\sim  90 \%$  at $\rm  K_T=19$.  At  $\rm  K_T=17.5$, the
typical  photometric accuracies on  the B-,  V-, R-  and I-  band data
amount to 0.38, 0.17, 0.11 and 0.1~mag, respectively.  For the optical
catalogues, we also  computed the fraction of sources  ($\rm f_S$) for
which  S-Extractor provides  reliable  magnitude estimates\footnote{We
adopted an upper limit of $100\%$  for the uncertainty on the flux.  }
as a  function of  the K-band  magnitude.  For the  B band,  $\rm f_S$
amounts to  $\sim 90\%$  at $\rm K_T=17.5$,  dropping to  $\sim0.5$ at
$\rm   K_T=19$.   For   the  V,   R  and   I  bands,   we   have  $\rm
f_S(K_T=17.5)=0.95,0.98,0.98$  and $  \rm f_S(K_T=19)=0.76,0.84,0.88$,
respectively.

\subsection{The catalogue}
\label{CATALOG}
\begin{figure*}
  \vspace{-4cm}
  \hspace{0cm}
  \includegraphics[width=13cm,height=28cm,angle=180]{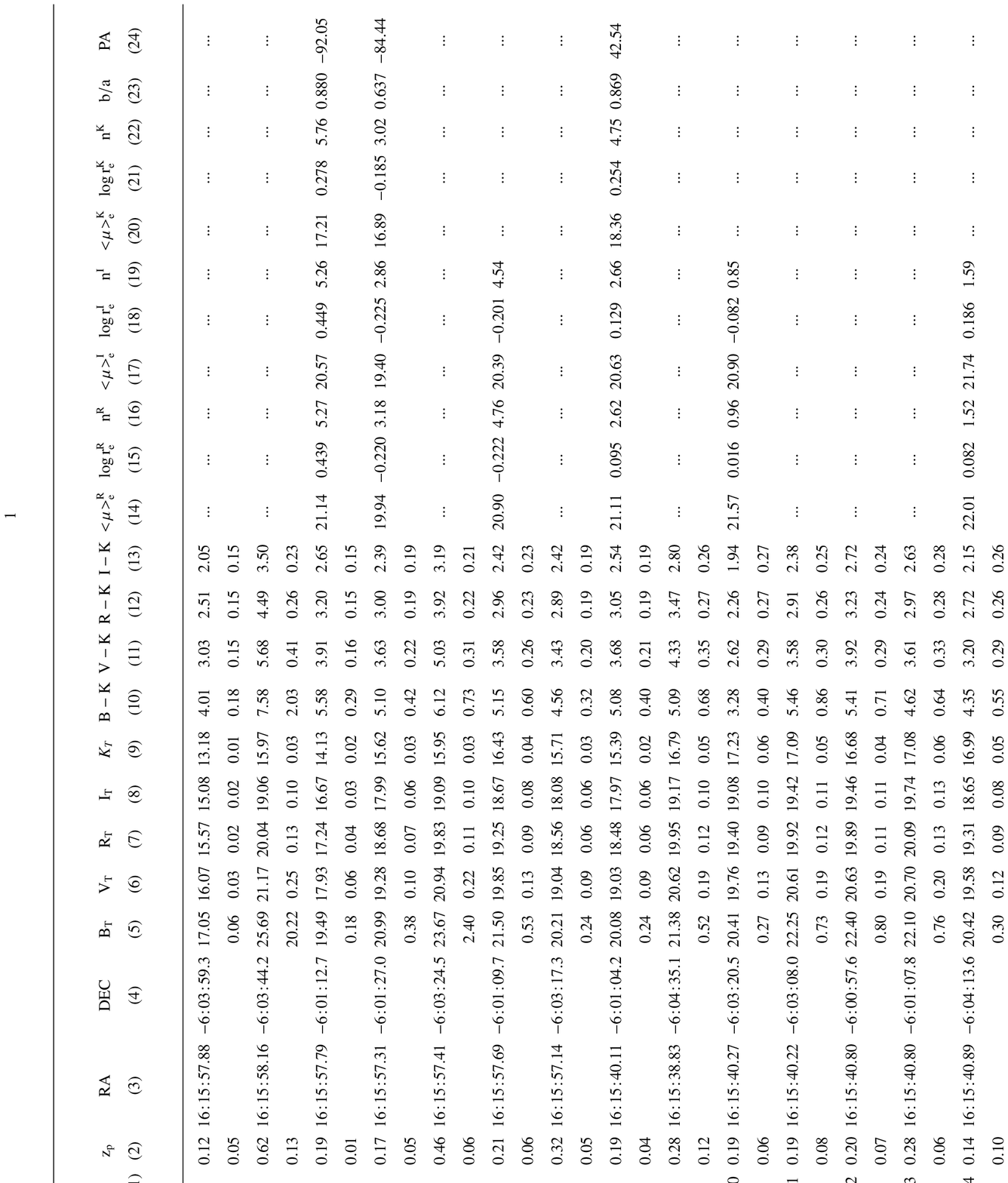}
  \label{CATA}
\end{figure*}
\begin{figure*}
  \centering 
  \includegraphics[angle=0,width=18cm,height=8.1cm]{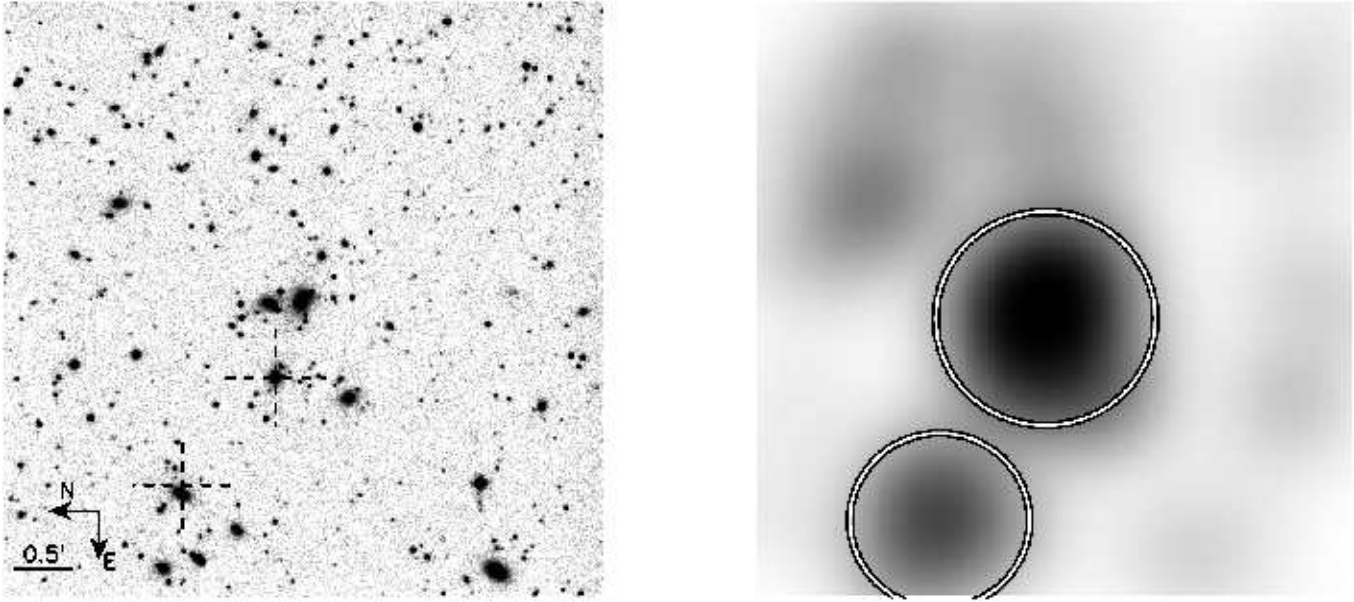}
  \caption{K-band  image  (left  panel)  and  luminosity-weighted  map
  (right  panel) for  the field  of  galaxies \A2163B.   The image  is
  centered at  $\rm RA= 16h \,  15m \, 48s$ and  $\rm DEC=-06^\circ \,
  02'  \,  10''  $, and  was  smoothed  for  the  plot by  a  gaussian
  convolution.  The scale  and the orientation of the  image are shown
  in the lower left corner of  the left panel.  In the same panel, the
  crosses indicates the  positions of the radio sources  in the field.
  In the right plot, the circles enclose the regions that were used to
  calculate the LF  in the higher density environment
  (see Sec.~\ref{LFKSEC}). }
  \label{CL16_K}
\end{figure*}

The  BVRIK photometric  catalogue for  the $\rm  N=131$  galaxies with
photometric redshift estimate (see Sec.~4)\footnote{The full catalogue
will be available  in electronic form at the CDS  via anonymous ftp to
cdsarc.u-strasbg.fr.}  is shown in Tab.~2.

The catalogue  is organized as  follows.  Two rows correspond  to each
galaxy.   The first one  provides the  measured quantities,  while the
second  one  gives  the  relative uncertainties  (one  sigma  standard
intervals).  All  the photometric  quantities have been  corrected for
galactic extinction  by using  the photometric redshift  estimates and
adopting $\rm E(B-V)=0.415$, with  the extinction law from SFD98.  The
typical uncertainties  on the  structural parameters are  discussed in
Sec.~\ref{PARSEC}.  \\ Column 1:  running number of the catalogue.  \\
Column 2: photometric  redshift.  \\ Column 3, 4:  right ascension and
declination referred to J2000.   The astrometric solution was computed
by using  a list  of stars from  the USNO  catalogue.  The rms  of the
residuals  to the  astrometric solution  is $\rm  0.2''$.   \\ Columns
5--9: B-, V-, R-, I-  and K-band total magnitudes.  \\ Columns 10--13:
optical--NIR galaxy  colours in the aperture of  $3.5''$ corrected for
seeing effects.   \\ Columns 14--16:  mean surface brightness  \mie \,
within the effective  radius \re, logarithm of \re  \, (in arcsec) and
Sersic index $\rm  n$ for the R-band.  \\ Columns  17--19: the same of
columns 14--16 for the I-band.  \\ Columns 20--22: the same of columns
14--16 for  the K-band.   \\ Columns 23,  24: axis ratio  and position
angle in degree obtained from the K-band surface photometry. \\

\section{The cluster \A2163B}
\label{THECLUSTER}
Fig.~\ref{CL16_K}  plots  the  K-band  image  of \A2163B  \,  and  the
luminosity-weighted density map for  the $\rm N=105$ galaxies brighter
than $\rm K_T=17$. The figure  clearly shows the presence of a cluster
of galaxies,  with a main over-density  in the center  and a secondary
structure at a distance of $\sim 2'$ ($\rm \sim0.4~Mpc$) from the main
peak.   We   note  that,   on  the  basis   of  field   galaxy  counts
(Sec.~\ref{LFKSEC}),  only  $\rm  \sim15$  galaxies  out  of  105  are
expected to  be field  contaminants in the  area of the  K-band image,
confirming that  the features  in Fig.~\ref{CL16_K} (right  panel) are
highly  significant.  It  is very  interesting to  note that  both the
X-ray image of \A2163B \, and  the $\rm20~cm$ radio map show a spatial
structure very similar to the  luminosity density image (see Fig.~7 of
FFG01).  In  Fig~\ref{CL16_K}, we also  show the positions of  the two
radio sources found  by FFG01. One of this turns  out to coincide with
an early-type\footnote{Sersic index greater than 2.}  galaxy (possible
merging) in  the main clump, while the  second one is very  close to a
bright early-type in the second overdensity region.

\begin{figure*}
  \centering
  \includegraphics[angle=0,width=12cm,height=12cm]{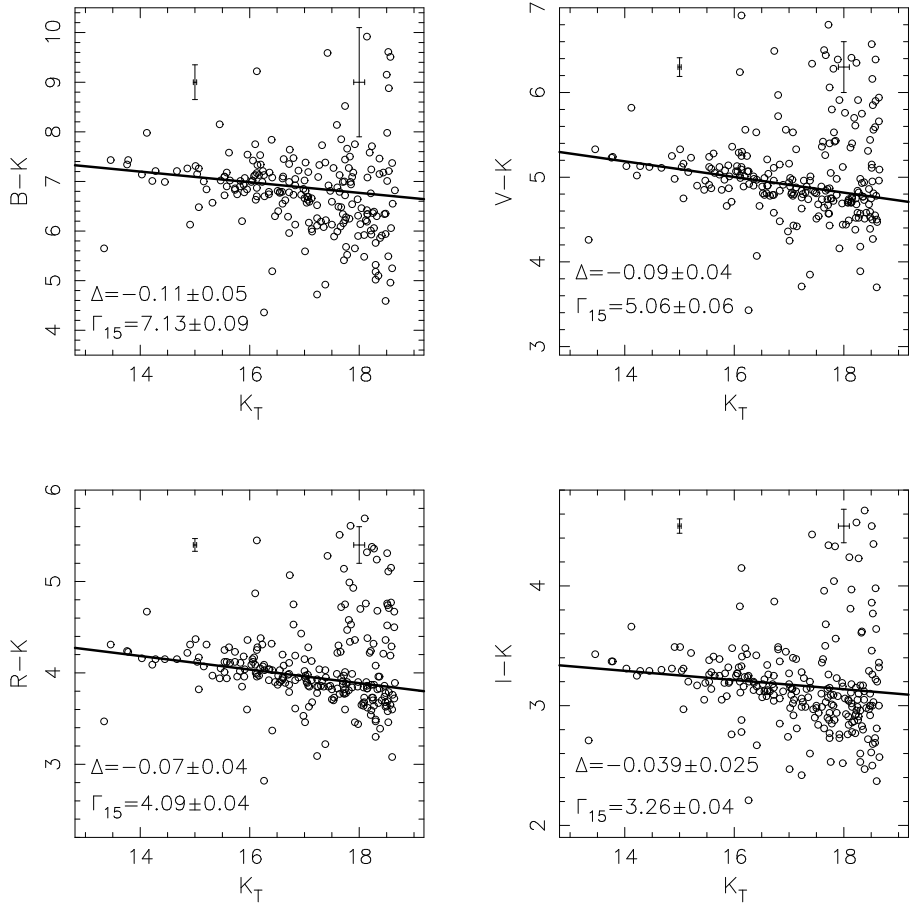}
  \caption{Optical--NIR  CM  diagrams for  galaxies  in  the field  of
 \A2163B. For each panel, the  slope and zeropoint of the CM relations
 are shown  in the lower  left corner, while  the error bars  from the
 left to the right denote  the averaged uncertainties on colours and
 magnitudes at $\rm K_T=15$ and $\rm K_T=18$, respectively.  }
  \label{CM_CL16}
\end{figure*}

In order to derive the redshift  $\rm z_c$ of the galaxy population of
\A2163B  \,  we compared  the  BVRIK  photometry  with predictions  of
stellar  population models  from the  GISSEL00 code  \citep{BrC93}. In
particular, we  used (1) the optical--NIR  colour magnitude (hereafter
CM)    relations    (Sec.~\ref{CMSEC}),     (2)    the    K-band    LF
(Sec.~\ref{LFKSEC})  and   (3)  the  photometric   redshift  technique
(Sec.~\ref{ZPSEC}).   We point  out that,  although it  is  known that
broad-band colours  are degenerate  in age, metallicity  and redshift,
and  that uncertainties  exist among  stellar population  models (e.g.
\citealt{CWB96}),  the  large  wavelength  baseline  provided  by  our
photometry  (from  U to  NIR  restframe) and  the  use  of the  GISSEL
synthesis  code  allow  to  obtain  a reliable  redshift  estimate  of
galaxies  at   intermediate  redshift  by   using  purely  photometric
techniques,  as   shown  by  other   studies  (e.g.   \citealt{dPP98},
\citealt{BML02},  \citealt{LMI03c}).    Another  important  point  for
estimating  $\rm z_c$  is the  effect of  the galactic  reddening $\rm
E(B-V)$.   The main sources  for the  reddening of  the Milky  Way are
based on $\rm HI$  measurements from \citet[hereafter BH84]{BuH84} and
on   COBE/DIRBE   and  IRAS/ISSA   FIR   data  from   \citet[hereafter
SFD98]{SFD98}.  Generally,  the estimates  of $\rm E(B-V)$  from SFD98
are systematically  higher than those  of BH84. For \A2163B,  BH84 and
SFD98 predict very different  extinction values, $\rm E(B-V)=0.18$ and
$\rm E(B-V)=0.415$,  respectively.  In the  next section, we  obtain a
simultaneous  estimate of  the galactic  reddening as  well as  of the
cluster  redshift from  the optical--NIR  colour  magnitude relations.
This  procedure  gives  an  extiction  value which  is  in  remarkable
agreement with that of SFD98 and  a redshift estimate of $\rm z_c \sim
0.2$, which concides within  the uncertainties with that obtained from
the K-band  LF (Sec.~\ref{LFKSEC}),  for which extinction  effects are
negligible.  The  difference between BH84 and  SFD98 reddening values,
therefore, does not affect the present analysis.

\subsection{Colour magnitude relations}
\label{CMSEC}
Fig.~\ref{CM_CL16} shows  the optical--NIR CM  diagrams, not corrected
for  galactic   extinction,  for   all  the  galaxies   brighter  than
$\mathrm{K_T=18.6}$  in the  field of  \A2163B. Each  diagram  shows a
sharp sequence of red galaxies, following the CM relation:
\begin{equation}
\mathrm{ N-K= \Gamma_{K_r}+ \Delta \cdot (K_T - K_r) },
\label{CMEQ}
\end{equation}
where  N is  the optical  waveband, $\rm  K_r$ is  a  reference K-band
magnitude, $\rm  \Delta$ is  the slope and  $\rm \Gamma_{K_r}$  is the
zeropoint of the  relation, that is the $\rm N-K$  colour at $\rm K_T=
K_r$.  The  values of  $\rm \Gamma_{K_r}$ and  $\rm \Delta$,  that are
shown in Fig.~\ref{CM_CL16}, were estimated by applying a least square
fit,   minimizing  the   rms  of   the  residuals   in   colour,  $\rm
\sigma_{N-K}$,  to the  relation.  In  order to  reduce the  effect of
outliers and  field contaminants, the value of  $\rm \sigma_{N-K}$ was
computed  by using the  bi-weight statistics  \citep{BFG90}. Moreover,
for comparison with previous works (KAB98), the fits were performed by
considering only the galaxies within the three brightest magnitudes of
the red sequence,  corresponding to $\rm K_T \lesssim  16.5$.  We note
that,  down to  this  limit,  the BVRIK  photometry  is complete  (see
Sec.~\ref{APSEC}), and that, on the  basis of field galaxy counts (see
Sec.~\ref{LFKSEC}), only  $\sim10 \%$ of the galaxies  are expected to
be field contaminants.  Despite the large uncertainties, the values of
$\rm  \Delta$  show  that  the  slope of  the  CM  relation  increases
systematically from about $ -0.1$  for the $\rm B-K$ sequence to about
$-0.04$ for the $\rm I-K$ relation.  The slopes of the $\rm V-K$, $\rm
R-K$ and  $\rm I-K$  sequences can be  straightly compared  with those
predicted from a pure mass-metallicity  model by KAB98 at the redshift
of \A2163B  ($\rm z \sim 0.2$, see  below): $\rm \Delta_{V-K}=-0.075$,
$\rm  \Delta_{R-K}=-0.062$, and  $\rm  \Delta_{I-K}=-0.05$ (see  their
Fig.~4). We note, in fact, that the CM slopes of KAB98 were normalized
to a physical aperture of $\rm \sim 10 ~kpc$, which is very similar to
that  adopted in  the present  work,  and, therefore,  the effects  of
galaxy colour  gradients do  not affect the  comparison of  CM slopes.
Our results fully agree with the values of KAB98.

The zeropoints of  the CM relation were used  to obtain a simultaneous
estimate  of  the cluster  redshift  $\rm  z_c$  and of  the  galactic
reddening  in the  direction of  \A2163,  fitting the  values of  $\rm
\Gamma_{K_r}$ with  the colours  expected from old  stellar population
models (see Appendix~B  for details).  In this approach,  we took into
account  the  degeneracy  among  metallicity  and  the  other  stellar
population parameters exploiting the  fact that the CM relation mainly
origins  from  a mass-metallicity  relation  (see \citealt{MLM03}  and
references therein).   This procedure gives $\rm  z_c=0.215 \pm 0.015$
and $\rm E(B-V) = 0.41 \pm  0.02$. It is quite remarkable that (1) the
estimate of $\rm E(B-V)$ is in very good agreement with that of SFD98,
$\rm E(B-V)=0.415$, and (2) the value of $\rm z_c$ coincides with that
obtained from the K-band LF, which is virtually unaffected by galactic
reddening.   For  this  reason,  we  adopted  in  the  following  $\rm
E(B-V)=0.415$, and applied the corresponding absorption corrections to
our photometry\footnote{For  $\rm E(B-V)=0.415$, at $\rm  z \sim 0.2$,
and  practically   for  any  galaxy  spectral  model,   we  have  $\rm
A_B=1.630$, $\rm A_V=1.281$, $\rm A_R=1.049$, $\rm A_I=0.788$ and $\rm
A_K=0.152$.}.  Repeating the zeropoint fitting with $\rm E(B-V)=0.415$
gives $\rm z_c \sim 0.21$, which is fully consistent with the redshift
of the main cluster structure (\A2163 ) at $\rm z \sim 0.2$.

\begin{figure}
  \centering
  \includegraphics[angle=0,width=8.5cm,height=8.5cm]{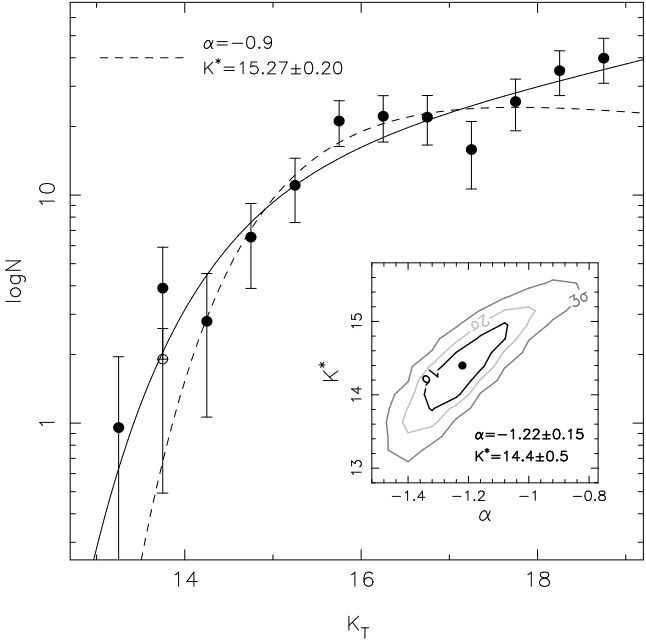}
  \caption{  K-band  LF  of  \A2163B.   The two  curves  indicate  the
   best--fit Schecter functions obtained  by fitting both $\alpha$ and
   $\rm K^*$ (solid line) and  by fitting only $\rm K^*$ with $\alpha$
   fixed  to  $-0.9$  (dashed  line).  In the  last  case,  the  three
   brightest cluster  galaxies were excluded  from the fit:  the first
   bin was not  considered while the galaxy counts  for the second bin
   are  marked by  the empty  circle.  Contours  in the  smaller panel
   denote  $1\sigma$  ,$2\sigma$  and  $3\sigma$  probability  levels,
   respectively.  }
  \label{LFK_CL16}
\end{figure}

\subsection{Luminosity function}
\label{LFKSEC}
The evolution of the characteristic  magnitude $\rm K^*$ of the K-band
LF  of cluster  galaxies  at $z  \lesssim  1$ is  consistent with  the
behaviour of an old passive stellar population with formation redshift
$\rm  z_f \in  [2,3]$ and  solar  metallicity.  This  result has  been
obtained  by   \citet[hereafter  dPS99]{dPS99}  fitting   a  Schechter
function (hereafter  SF) with fixed slope $\alpha=-0.9$  to the K-band
counts of cluster galaxies with $\rm K_T-K^* < 3$, and turns out to be
independent of cluster richness or X--ray luminosity.

We used  these results in  order to obtain  a further estimate  of the
redshift of  \A2163B.  The  K-band LF was  obtained by  correcting our
number  counts with  field  galaxy  counts from  the  Calar Alto  Deep
Imaging  Survey (CADIS,  see  \citealt{HTK01}), which  is the  largest
medium  deep  K-band  survey  to  date,  with a  total  area  of  $\rm
0.2~deg^2$ and a completeness  magnitude of $\rm 19.75~mag$. Since our
K-band   photometry   is   complete   down  to   $\rm   K_T=19$   (see
Sec.~\ref{APSEC}), the  CADIS data are  the most suitable  dataset for
the estimate of field galaxy counts.  The LF of \A2163B \, is shown in
Fig.~\ref{LFK_CL16}.   The  error bars  take  into account  Poissonian
uncertainties on  both field and  cluster counts, while  the different
curves show the SF fits of the  LF. In order to account for the finite
size of the magnitude bins, the fits were performed by the convolution
of  a SF  with the  bin  size.  Following  dPS99 and  \citet[hereafter
dPP98]{dPP98}, we fitted the galaxy counts down to $\rm K_T=18.$ ($\rm
\sim  K^*+3$, see  below) by  a SF  with $\alpha=-0.9$,  excluding the
three brightest cluster galaxies.  The fit provided a good description
of the counts of bright galaxies\footnote{ The only two bins for which
galaxy counts  deviate by  more than $1\sigma$  from the model  are at
$\rm K_T > 17$. We note that  limiting the fit range to $\rm K_T < 17$
does  not affect  our results,  giving $K^*_{0.9}=  15.34  \pm 0.2$.},
giving the  following estimate  of the characteristic  magnitude: $\rm
K^*_{0.9}  = 15.3  \pm  0.20$.  In  order  to estimate  $\rm z_c$,  we
minimized the function:
\begin{equation}
\mathrm{
 \chi^2= \left[ K^*_{0.9}(z)- K^*_{0.9} \right]^2,
}
\end{equation}
 where  $\rm K^*_{0.9}(z)$  is  the  value of  $\rm  K^*$ expected  at
redshift  $\rm z$,  and was  obtained by  adding to  the value  of the
characteristic  magnitude of  the Coma  cluster ($10.9  \pm  0.2$, see
dPP98) the  luminosity distance term and the  K+E corrections, derived
from  GISSEL00 spectral  models with  a Scalo  IMF and  an exponential
SFR. It turned out that,  by using models with different metallicities
($\rm Z/Z_\odot = 0.5, 1.,  2.$) and formation redshifts ($\rm z_f \ge
1.5$), the estimate of $\rm z_c$ does not vary significantly: $\rm z_c
= 0.20  \pm 0.02$.  This  result is in  very good agreement  with that
obtained from  the CM  relations in Sec.~\ref{CMSEC},  and, therefore,
still shows  that the redshift of  \A2163B is consistent  with that of
the main cluster structure \A2163.  We found that the estimate of $\rm
z_c$ is  very robust with  respect to the field  subtraction procedure
and to  the value adopted for  the galactic reddening, due  to the low
sensitivity of the K-band photometry to dust absorption.  For example,
by using $\rm  E(B-V)=0$, we obtain $\rm z_c  = 0.225$, and neglecting
field counts gives $\rm z_c = 0.21$.

\begin{figure}
  \centering
  \includegraphics[angle=0,width=8.5cm,height=8.5cm]{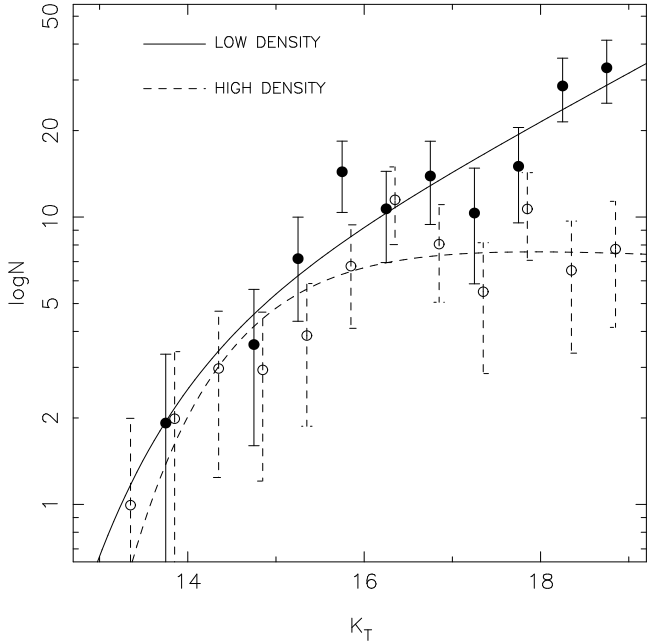}
  \caption{  K-band  LF of  \A2163B  in  the  low- and  high-  density
   environments.  The  points and the error bars  corresponding to the
   high density region have been shifted by $\rm +0.1~mag$ for sake of
   semplicity.  The curves show the best fitting Schechter functions.}
  \label{LF_K2}
\end{figure}

In order to study  the faint end of the LF, we  performed a SF fit for
$\rm  K_T <  19$  by treating  both  $\alpha$ and  $\rm  K^*$ as  free
parameters.  As  shown in Fig.~\ref{LFK_CL16}, the  SF, whose best-fit
parameters are $\alpha=-1.22 \pm 0.15$ and $\rm K^* = 14.45 \pm 0.50$,
provides a good description of  the LF of \A2163B.  The cluster counts
show some  deviation from  the SF fit  only for  the two bins  at $\rm
K_T=15.75$ and $\rm K_T=17.25$.   Although this could be an indication
of a  bimodal behaviour of the  LF, as found by  previous studies (see
\citealt{MMM03}, and  references therein), the deviations  are only at
the  levels  of $1.6\sigma$  and  $1.7\sigma$,  respectively, and  are
therefore not very significant.  The value of $\alpha$ is in very good
agreement with the value of  $-1.18$ found by \citet{AND01} for a rich
cluster of galaxies at $\rm z=0.3$ in an area of $\rm \sim 1.3~Mpc^2$,
which  is similar to  that analyzed  in the  present study  ($\rm \sim
1~Mpc^2$), and is  also consistent with values of  the faint end slope
of  the  NIR  LF  for  the Coma  cluster,  $\alpha=-1.0$  (dPP98)  and
$\alpha=-1.3$  \citep{AnP00}, and  for field  galaxies, $\alpha=-1.09$
\citep{KPF01}.  We also  analyzed the dependence of the  LF of \A2163B
\,  on environment,  by  performing  the SF  fits  separately for  the
galaxies inside the  higher density region, defined by  the circles in
Fig.~\ref{CL16_K}, and for the  galaxies outside that region.  The two
LFs  are shown  in Fig.~\ref{LF_K2}  with the  corresponding  SF fits.
Interestingly,  we  found  that  the  values  of  $\alpha$  vary  from
$-0.96\pm0.18$  (higher  density)  to $-1.42\pm0.2$  (lower  density),
implying a  steepening of the faint  end slope at  lower densities, in
agreement with the findings of  other studies (see Haines et al.~2003,
submitted, and references therein). This steepening is due to the fact
that, as  shown in Fig.~\ref{LF_K2}, the lower  density environment is
characterized  by  a  higher   fraction  of  faint  galaxies,  with  a
significant difference  ($> 2 \sigma$)  in the low- and  high- density
counts at $\rm K_T > 18$.

\section{Photometric redshifts}
\label{ZPSEC}
In order to  select cluster members, we used  the photometric redshift
technique, which has  been proven to be an  effective tool to estimate
the  redshift  of cluster  galaxies  at  intermediate redshifts  (e.g.
\citealt{BML02},    \citealt{LMI03c}    and    references    therein).
Photometric redshifts were estimated  according to the Spectral Energy
Distribution  fitting method (see  \citealt{MAa, MAb},  and references
therein).  In  order to achieve  a reasonable accuracy,  we considered
galaxies with  signal-to-noise ratio $\rm S/N  > 5$ in  at least three
bands, limiting the  sample to the $\rm N=131$  galaxies brighter than
$\rm  K_T=17.5$.   Since   the  percentage  of  galaxies\footnote{This
estimate was obtained by using  a Pure Luminosity Evolution model (see
e.g.  \citealt{POZ}).}  at redshift $\rm z > 1$ with $\rm K < 17.5$ is
expected to be  negligible ($< 2 \%$), we looked  for redshifts in the
range $z \in [0.0,1.0]$ with a  step of 0.01, imposing that at a given
redshift galaxy templates were younger than the age of the universe in
the adopted cosmology.  We used  the GISSEL00 code in order to produce
galaxy  templates  with  a  Scalo  IMF and  an  exponential  SFR  $\rm
e^{-t/\tau}$.   The colours  of E/S0,  Sa/Sb, and  Sc/Sd  spectra were
modeled by choosing  $\rm \tau = 1,4$ and  $\rm 15~Gyr$, respectively,
while early-type galaxies  with different metallicities were described
by  using E/S0  models  with  $Z/Z_\odot$=0.2, 0.4,  1  and 2.5.   The
differential  dust extinction  of the  Milky Way  was included  in the
computation  of model  colours  by adopting  the  extinction curve  of
\citet{CAR}   and  a   colour  excess   of  $\rm   E(B-V)=0.415$  (see
Sec.~\ref{CMSEC}).  The  uncertainty on the  photometric redshift $\rm
\delta z$  was estimated  by performing numerical  simulations, taking
into account the measurement errors on galaxy colours.  The mean value
of $\delta  z$ is $\sim0.08$, varying  from $\sim 0.05$ at  $\rm K_T =
15$ to $\sim 0.1$ at $\rm K_T = 17$.

\begin{figure}
  \centering 
  \includegraphics[angle=0,width=8cm,height=8cm]{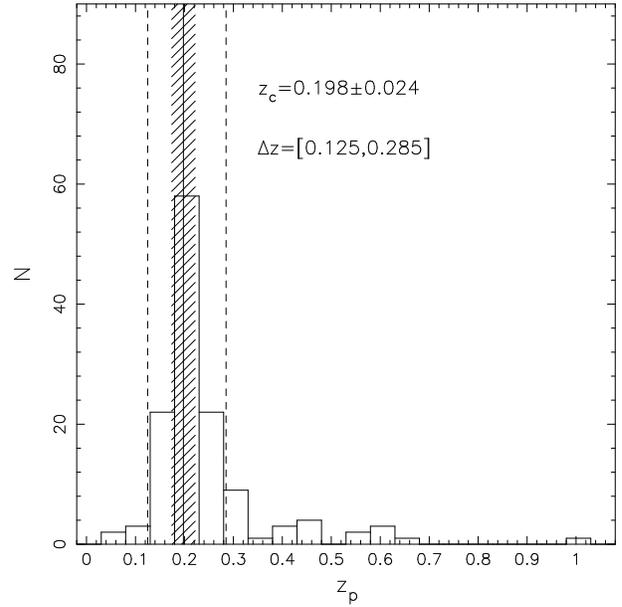}
  \caption{ Distribution of photometric  redshifts. The solid line and
   the dashed  region indicate the  estimate of the  cluster redshift,
   $\rm  z_c$,  and   the  relative  uncertainty  ($1\sigma$  standard
   interval). Dashed lines enclose the range of $\rm z_p$, $\Delta z$,
   used to select cluster members.  }
  \label{ZP_CL16}
\end{figure}

  As  shown  in Fig.~\ref{ZP_CL16},  the  distribution of  photometric
redshifts is  dominated by  the peak around  $z \sim  0.2$, indicating
that most galaxies with $\rm K_T < 17.5$ are actually cluster members.
Applying the bi-weight statistics,  we obtained the following estimate
of the  cluster redshift:  $\rm z_c =  0.198 \pm 0.024$,  in agreement
with   the  values   obtained  in   Sec.\ref{CMSEC}  and~\ref{LFKSEC}.
Galaxies with  photometric redshift in a range  of $2 \overline{\delta
z}$ around $\rm  z_c$ were defined as cluster  members, resulting in a
final  list  of  $\rm  N=102$  objects\footnote{We note  that  in  the
analysis of the optical-NIR colour gradients we consider only the $\rm
N  \sim 60$ galaxies  brighter than  $\rm K_T  = 16.5$.   Proceding as
described  in  \citet{LMI03c},  we  estimated that  at  this  limiting
magnitude, in  the redshift range  adopted to select  cluster members,
only $\sim 1\%$ of the galaxies are expected to be field contaminants,
proving  the reliability  of  the cluster  members  selection for  the
surface photometry  analysis.}.  This value is in  good agreement with
that  predicted from  the LF  of  cluster galaxies,  $\rm N=112$  (see
Sec.~\ref{LFKSEC}).

\begin{figure*}
  \centering 
  \includegraphics[angle=0,width=14cm,height=12cm]{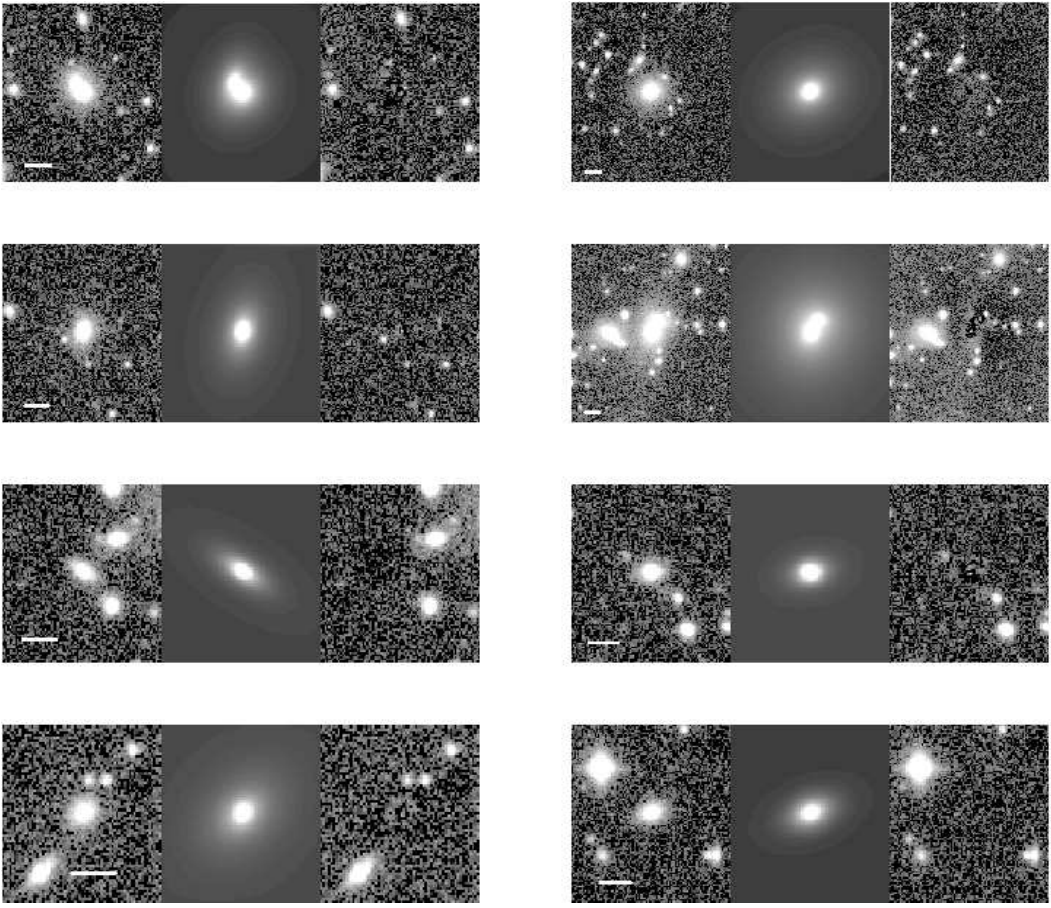}
  \caption{ 2D fits of galaxies in the K-band image. For each plot,
  from left to right, the galaxy image, the 2D model and the residual
  map are shown. The size of the horizontal white bar is $5``$.  }
  \label{2DFIT}
\end{figure*}

\section{Surface photometry}
\label{SPSEC}

Surface photometry was derived for all the galaxies brighter than $\rm
R_T = 19.5$ in  the R and I bands, and for  the galaxies brighter than
$\rm K_T =  16.5$ in the K band. These  selection criteria were chosen
in order to obtain reliable  structural parameters from the \A2163B \,
photometry, and were the results of numerical simulations performed as
described  in  LBM02.   A  further  selection  was  performed  {\it  a
posteriori}  by excluding  few faint  galaxies in  the I  and  K bands
because of the strong  contamination from closed stars.  The resulting
samples include  $\rm N=64, 62$ and $58$  galaxies for the R,  I and K
bands, respectively, with  $\rm N=62$ objects in common  between the R
and I  bands, and $\rm  N=58$ galaxies in  common between the R  and K
bands.   Structural parameters  for these  galaxies are  given  in the
catalogue (see Sec.~2.3).

Galaxy images  were fitted with 2D  models convolved with  the PSF, as
detailed in  LBM02.  As  shown in that  paper, the 2D  approach allows
reliable  structural  parameters  to   be  obtained  for  galaxies  at
intermediate redshift ($\rm z \sim  0.3$) from ground based data taken
under ordinary observing conditions  (seeing $\rm FWHM \sim 1.0''$ and
pixel scale $\rm \sim 0.3''/pxl$).
Galaxy models were parametrized by the Sersic law:
\begin{equation}
\mathrm{
I(r) = I_0 \cdot exp(- b \cdot (r / R_e)^{1/n}),
}
\end{equation}
where  r is  the equivalent  radius, $\mathrm{R_e}$  is  the effective
(half-light) radius, $\mathrm{I_0}$ is the central surface brightness,
$\mathrm{n}$  is  the Sersic  index,  and b  is  a  constant ($b  \sim
2n-1/3$, see  \citealt{CCD93}).  For each galaxy,  nearby objects were
masked   interactively,  while   overlapping   galaxies  were   fitted
simultaneously.   The  $\chi^2$  minimization  was  performed  by  the
Levenberg-Marquardt   method,  deriving  a   total  of   six  best-fit
parameters: the center coordinates,  the effective radius, the central
surface brightness,  the position  angle $\rm PA$  and the  axis ratio
$\rm b/a$. The mean surface brightness within \re, \mie, and the total
magnitude \mtot \, were computed  from the fitting parameters by using
the well known properties of  the Sersic law (see \citealt{CB99}). The
typical uncertainties on galaxy parameters were estimated by comparing
structural parameters between  different wavebands (see next section).
The  PSF models  were obtained  by a  multi-Gaussian expansion  of the
images of  bright unsaturated stars in  the cluster field.   For the R
and I bands, all the  stars were fitted simultaneously, resulting in a
single PSF  model. For the K band,  the PSF turned out  to vary across
the field,  showing significant deviations from the  circular shape in
the North-East corner  of the frame. Since star  images were available
at  different  positions across  the  field,  with  an almost  uniform
coverage, we adopted  for each galaxy the PSF  model obtained from the
closest star.  PSF distortions were  treated as in LBM03a, by adopting
gaussian  functions with  elliptical  isophotes.  \\  Fig.~\ref{2DFIT}
plots, as  an example, the K-band  2D fitting of some  galaxies in the
cluster field. The  models turned out to give  a very good description
of the galaxy images.
\begin{figure*}
  \centering
  \includegraphics[angle=0,width=10cm,height=10cm]{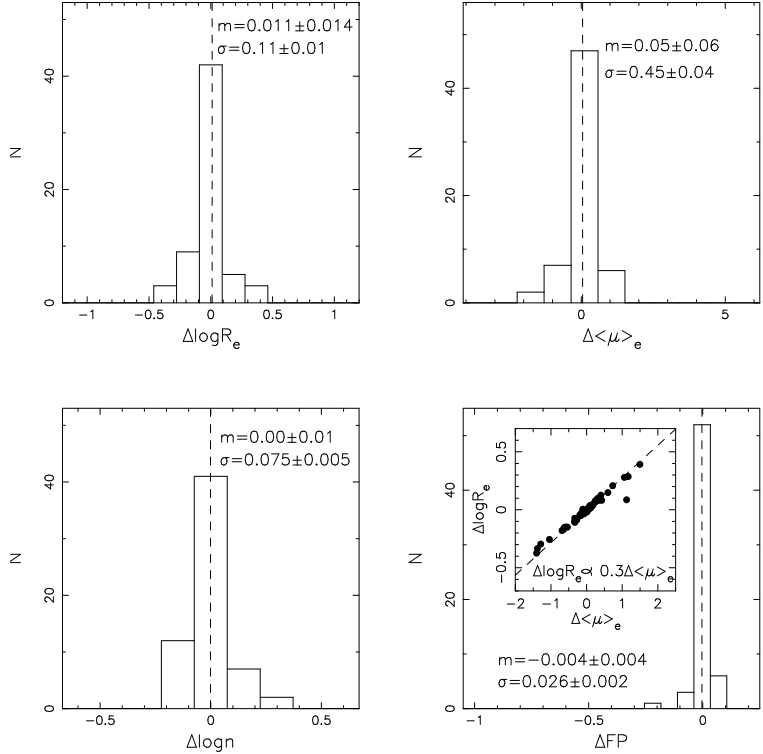}
  \caption{Comparison  of structural  parameter in  the  optical.  For
   each panels,  differences are computed  between R and I  bands. The
   correlation between effective parameters  is shown inside the lower
   right panel. }
  \label{SP_RI}
\end{figure*}

\section{Optical--NIR internal structure of cluster galaxies}
The   waveband  dependence  of   the  structural   parameters  carries
information  on the  differential properties  of SPs  inside galaxies.
This  subject  was investigated  in  LBM03a,  who  first attempted  to
analyze the evolution of the UV--NIR structural properties of both the
populations of  disk dominated (Sersic  index $\rm n$ smaller  than 2)
and  spheroidal galaxies  up to  redshift $\rm  z\sim0.6$, by  using a
large sample of cluster  galaxies ($\rm N=270$).  Following that work,
we studied the internal optical--NIR structure of galaxies for \A2163B
by    (1)   a   straight    comparison   of    structural   parameters
(Sec.~\ref{PARSEC}) and  (2) estimating the  internal colour gradients
of galaxies  (Sec.~\ref{CGRAD}). Since the  number of disks  in common
between the R  and K bands turned out to be  negligible ($\rm N=5$ out
of 58), we did not  attempt\footnote{Disk galaxies are included in the
analysis of  Sec.~\ref{PARSEC} and~\ref{CGRAD}.  However,  the results
are unchanged by excluding these galaxies from the samples.}  to study
the waveband dependence of their structural properties.  The following
results  will  refer,  therefore,  to  the  spheroidal  population  of
\A2163B.
\begin{figure*}
  \centering
  \includegraphics[angle=0,width=10cm,height=10cm]{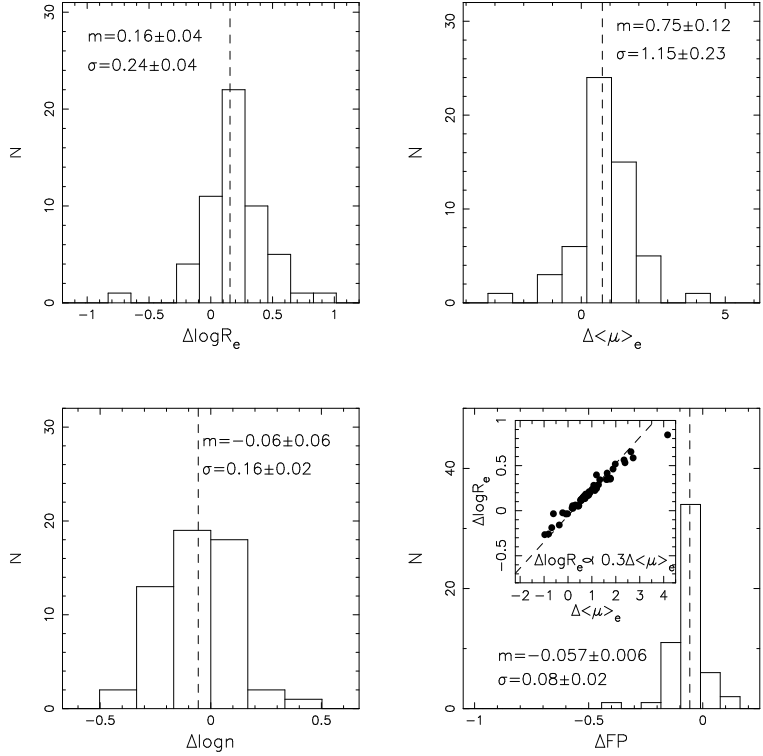}
  \caption{ Same of Fig.~\ref{SP_RI} for the R- and K-band parameters.
  }
  \label{SP_RK}
\end{figure*}

\subsection{Comparison of structural parameters}
\label{PARSEC}
Fig.~\ref{SP_RI} shows the comparison of structural parameters for the
$\rm N=62$ galaxies in common between the R and I bands. { Differences
are  always   computed  by   subtracting  the  quantities   at  longer
wavelengths  from  those   at  shorter  wavelengths}.   The  following
quantities   are  considered:   effective  radii   and   mean  surface
brightnesses, Sersic indices,  and the combination of \re  \, and \mie
\, which  enters the  Fundamental Plane: $\rm  k_{FP}= \log R_e  - 0.3
\cdot  < \!   \mu \!   >_e$.   We note  that the  differences of  mean
surface  brightnesses,  $\rm \Delta  \!   <  \!   \mu \!   >_e$,  were
obtained by  subtracting the $\rm  R-I$ galaxy colours, that  is: $\rm
\Delta \!   < \!  \mu  \!  >_e = <  \!  \mu \!   >_e^R - < \!   \mu \!
>_e^I - (R - I)$. The mean value, $\rm m$, and the standard deviation,
$\sigma$, of each  distribution are shown in the  figure, while in the
lower right panel we also  plot the correlation between $\rm \Delta \!
\log R_e$ and  $\rm \Delta \!  < \!  \mu  \!  >_e$. Optical parameters
are  in remarkable  agreement, with  mean differences  that  are fully
consistent with zero  for each quantity.  This is  consistent with the
fact  that  the  R  and  I  bands  at  $\rm  z  \sim  0.2$  correspond
approximately  to  V and  R  restframe,  sampling,  therefore, a  very
similar spectral region for  early-type galaxies.  For this reason, we
used   the   distributions  of   Fig.~\ref{SP_RI}   to  estimate   the
uncertainties on \lre,  \mie \, and \lns, by  computing the covariance
matrix of the  differences between the R- and  I-band parameters.  The
standard deviations of the distributions amount to $\sim 25\%$ in \re,
$\sim 0.45~\rm mag/arcsec^2$ for \mie \, and $\sim 18 \%$ for $\rm n$.
We note, however, that due to the well known tight correlation between
the   uncertainties   on    the   effective   parameters   (see   e.g.
\citealt{JFK95a}),  the  $\rm  k_{FP}$  values  have  a  much  smaller
dispersion, only $6 \%$.


Differences between  R- and K-band structural parameters  are shown in
Fig.~\ref{SP_RK}.   We  note  that  the differences  of  mean  surface
brightnesses, $\rm  \Delta \!   < \!  \mu  \!  >_e$, were  obtained by
subtracting the $\rm  R-K$ galaxy colours, that is:  $\rm \Delta \!  <
\!  \mu \!  >_e = < \! \mu \!  >_e^R - < \!  \mu \!  >_e^K - (R - K)$.
The mean  values of the  distributions show that cluster  galaxies are
more concentrated in the NIR than in the optical, having NIR effective
radii  which are on  average $\sim  40\%$ smaller  than in  the R-band
($\rm \overline{\Delta \log  R_e}=-0.16$), and Sersic indices slightly
larger  in  the K  band,  although the  last  difference  is not  very
significant.   The mean difference  of $\rm  \Delta \!   < \!   \mu \!
>_e$ can be fully explained by using the definition of total magnitude
$\rm m_T  = -2.5 \log( 2  \pi) -5 \log R_e  + < \!  \mu  \!  >_e$, and
computing the difference between R- and K-band total magnitudes:
\begin{equation}
 \mathrm{ R_T - K_T = -5 \Delta \log R_e + < \! \mu \! >_e^R - < \!
\mu \! >_e^K}.
\label{PARVAREQ}
\end{equation}
Since $\rm  \Delta < \! \mu  \! >_e \simeq <  \! \mu \!  >_e^R  - < \!
\mu \!  >_e^K - (R_T - K_T)$,  we obtain the relation $\Delta < \! \mu
\!  >_e =  5 \Delta \log R_e$, which, for  $\rm \overline{ \Delta \log
R_e}=-0.16$, gives $\overline{  \Delta < \!  \mu \! >_e  } \sim 0.8 $,
in   very   good   agreement    with   the   mean   value   shown   in
Fig.~\ref{SP_RK}. The same procedure allows the mean difference of the
$\rm k_{FP}$ variable to be explained. \\
The dispersion of the distributions  in Fig.~\ref{SP_RK} is due (1) to
the uncertainties on  both the R- and K-band  parameters, and (2) to
the intrinsic  scatter of the optical--NIR properties  of galaxies. We
verified by  numerical simulations  (performed as described  in LBM02)
that  the uncertainties  on the  K-band structural  parameters  are as
large as  or smaller than  those estimated for the  optical wavebands.
This   suggests   that   the   dispersions  of   the   histograms   in
Fig.~\ref{SP_RK} have mainly an intrinsic origin. In Secs.~\ref{KRSEC}
and~\ref{PPSEC}, in order to  describe the uncertainties on \lre, \mie
\,  and \lns,  we  adopted the  same  covariance matrix  for both  the
optical and the  NIR data.  By looking at  Fig.~\ref{SP_RK}, we stress
again the small  dispersion in the $\rm k_{FP}$ variable,  that amounts to
$\rm \sim 0.08~dex$.

\subsection{Colour gradients}
\label{CGRAD}
The  internal colour  gradient of  galaxies can  be computed  from the
values  of  the effective  radii  and of  the  Sersic  indices in  the
different wavebands, as described in LBM02 (see their Eq.~4).

\begin{figure*}
  \centering
  \includegraphics[angle=0,width=12cm,height=6cm]{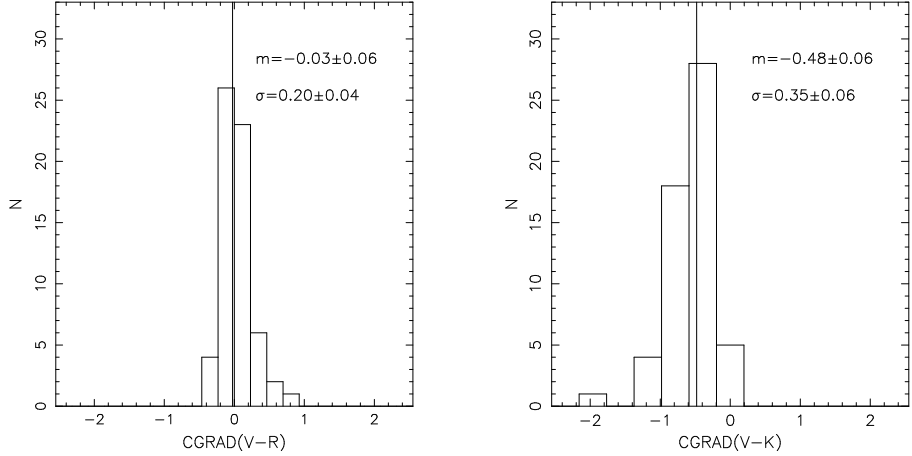}
  \caption{ Distributions of $\rm V-R$ and $\rm V-K$ colour gradients.
   The mean values are marked by a solid line.  }
  \label{CG}
\end{figure*}

The distributions of  colour gradients for the galaxies  of \A2163B \,
are  shown in  Fig.~\ref{CG}, we  wrote the  $\rm R-I$  and  $\rm R-K$
colour  gradients as  $\rm  V-R$ and  $\rm  V-K$ restframe  gradients,
respectively. By  using different  galaxy templates from  the GISSEL00
code, we verified,  in fact, that the conversion  from $\rm R-I$ ($\rm
R-K$)  at $\rm  z=0.2$ into  $\rm  V-R$ ($\rm  V-K$) at  $\rm z=0$  is
independent  of the spectral  type (for  $\rm z_f>1$),  with $1\sigma$
variations  of $0.01$  and $\rm  0.03~mag$ between  different spectral
models  for   the  optical--optical  and   the  optical--NIR  colours,
respectively.  As expected on  the basis of the structural parameters,
galaxies  show  optical--optical  colour  gradients  which  are  fully
consistent with zero, while the  V-K colour gradients are negative for
most galaxies,  implying that their SPs  are on average  redder in the
center. The mean value of  the optical--NIR gradient is $-0.48$, which
implies that  galaxies become $\rm  \sim 0.5~mag$ bluer per  decade of
radius toward  the periphery.  The  value of $\rm grad(V-K)$  is fully
consistent  with   the  colour  gradient  estimate   of  LBM03a  ($\rm
\sim~-0.4~mag/dex$).

\section{Optical--NIR Kormendy relations}
\label{KRSEC}
\begin{figure*}
  \centering
  \includegraphics[angle=0,width=13cm,height=6.5cm]{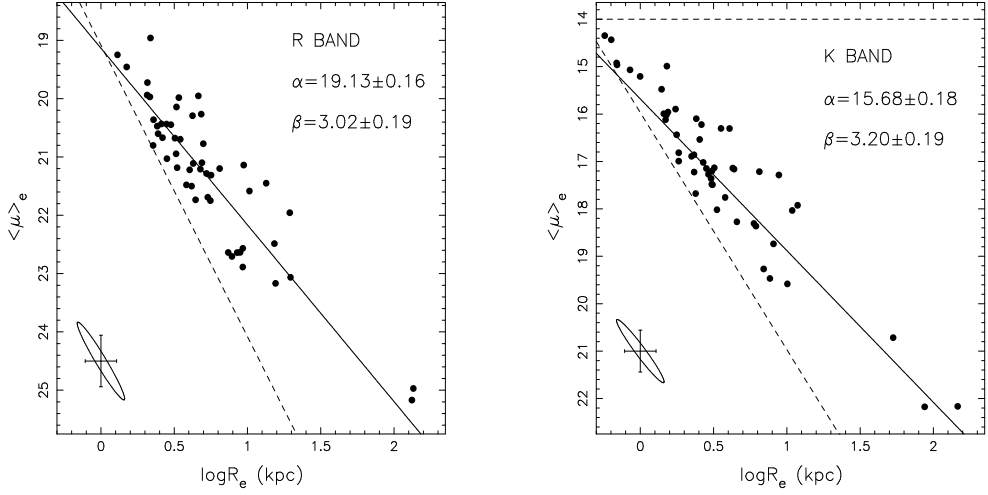}
  \caption{ Kormendy relations in R and K bands. Dashed lines mark the
  completeness magnitudes and the surface brightness cut (right panel)
  of  each   sample.  The  typical  uncertainties   on  the  effective
  parameters are indicated by the ellipses in the lower left corner of
  each panel. The ellipses mark the contours that enclose a probability
  of $68\%$ for a 2D normal deviate.  }
  \label{KRPLOT}
\end{figure*}

The \lre--\mie \, diagrams for the galaxies of \A2163B \, are shown in
Fig.~\ref{KRPLOT} for the R and K bands.  We note that the I-band data
provide almost the same spectral information as the R-band photometry,
and therefore are  not discussed in the following.   For each band, we
excluded the disks ($\rm n <  2$) and the galaxies for which the total
magnitude estimated  by the structural parameters is  fainter than the
completeness magnitude.  This selection results in $\rm N=52$ and $\rm
N=53$ galaxies in the R  and K bands, respectively.  For what concerns
the NIR data, we also excluded  two galaxies that have small radii and
low surface brightness and for which, therefore, structural parameters
are  possibly affected  by a  larger uncertainty.   The  K-band sample
consists of $\rm N=51$ objects.
In  order to  describe the  \lre--\mie  \, sequence,  we consider  the
following equation for each waveband $\rm M$:
\begin{equation}
\mathrm{
 < \! \mu \! >^M_e = \alpha_M + \beta_M \cdot \log R^M_e,
}
\label{KREQ}
\end{equation}
 where $\rm R_e$ is the  effective radius in $\rm kpc$, $\rm \alpha_M$
and $\rm  \beta_M$ are  the zeropoint and  the slope of  the relation.
The  evolution of the  slope, of  the zeropoint  and of  the intrinsic
dispersion  of the  KR has  been recently  studied in  the  optical by
LBM03b, who  showed that these properties do  not change significantly
over redshift at least up to $\rm z \sim 0.6$.  They found $\beta=2.92
\pm 0.08$, an intrinsic dispersion of $\rm 0.4 \pm 0.03~mag/arcsec^2$,
and a KR zeropoint of $\rm 18.95\pm0.08~mag/arcsec^2$ in the R band at
$\rm  z\sim 0.21$.   In order  to derive  the values  of  $\alpha$ and
$\beta$ for \A2163B, we applied  the $\rm BMLS$ fitting procedure (see
LBM03b), which allows the selection criteria to be taken into account,
providing  un-biased estimates  of the  KR coefficients.   The fitting
results  are  shown  in  Fig.~\ref{KRPLOT}.   For  what  concerns  the
dispersion of  the KRs, its value  turns out to  be consistent between
the  R and  K bands,  amounting  to $0.65\pm0.12$  in \mie  \, and  to
$0.22\pm0.04$ in \lre.  The R-band slope of the \lre--\mie \, relation
is  fully consistent  with that  of  LBM03b, in  agreement with  their
finding  that  this coefficient  does  not  evolve significantly  over
redshift.    By  adopting   $\beta=2.92$   in  the   fit,  we   obtain
$\alpha_R=19.16  \pm 0.1$  and, therefore,  also the  R-band zeropoint
turns out to be in full agreement with the result of LBM03b, once that
measurement  errors  are  taken   into  account.   By  subtracting  in
quadrature  from the  observed scatter  around  the KR  the amount  of
dispersion due to  the measurement uncertainties\footnote{To this aim,
we took into  account the correlation of the  uncertainties on \lre \,
and \mie.}  on  \lre \, and \mie, we obtain  the following estimate of
the  intrinsic  dispersion of  the  KR: $\rm  0.6\pm0.13~mag/arcsec^2$
($\rm 0.21\pm0.04~dex$ in  \re).  This value is a  little larger than,
but consistent with the estimate of LBM03b.

\begin{figure*}
  \centering
  \includegraphics[angle=0,width=13cm,height=6.5cm]{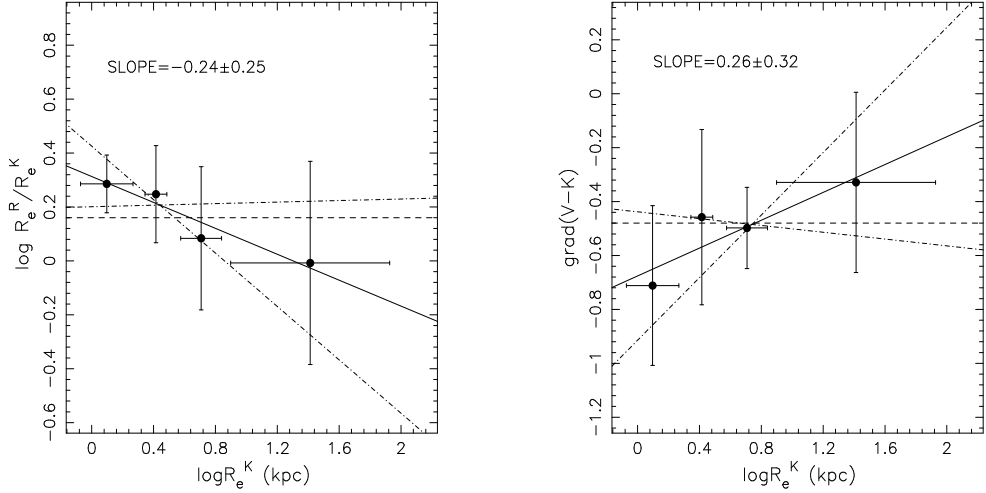}
  \caption{  Ratios  of optical  to  NIR  effective  radii and  colour
  gradients are  plotted versus the logarithm of  the K-band effective
  radius.  The dashed lines mark the  mean values of $\rm \log R_e^R /
  R_e^K$  and   $\rm  grad(V-K)$  in   the  left  and   right  panels,
  respectively. Data were  binned in order to have  the same number of
  points in each bin.  The solid lines mark the best-fits to the data,
  while  the  dot--dashed  lines  indicate  the  $1\sigma$  confidence
  intervals. }
  \label{CORPAR}
\end{figure*}

In order to compare the coefficients of the optical and NIR \lre--\mie
\, sequences,  we have to take  into account (1) the  variation of the
galaxy colour along the sequence  due to the colour magnitude relation
and (2) the waveband dependence of  \re \, and \mie \, due to internal
colour  gradients   of  galaxies.   By   using  Eqs.~1,~\ref{PARVAREQ}
and~\ref{KREQ}, we obtain the following relation:
\begin{equation}
\begin{array}{lll}
\mathrm{ < \!  \mu \! >^K_e } &= \mathrm{< \! \mu \! >^R_e  - 5 \cdot
 (\log R^R_e - \log R^K_e)- (R_T-K_T) } \simeq & \nonumber \\
  & \mathrm{ \left( 1 + \Delta \right)^{-1} \! \cdot  
 \left[ \alpha_R - \Gamma + 2.5 \cdot \Delta \cdot \log 2 \pi  - 
 \left( 5 - \beta_R \right) \cdot 
   \right. } & \nonumber \\
  & \mathrm{ \cdot \log \frac{R^R_e}{R^K_e} + \left. \left( \beta_R + 5 \Delta \right) \cdot \log
 R^K_e \right] } = & \nonumber \\
 & \mathrm{
  \alpha'_K + \beta'_K \cdot \log R^K_e + \gamma'_K \cdot \left( \log \frac{R^R_e}{R^K_e} 
 - \overline{ \log \frac{R^R_e}{R^K_e} } \right), } & \\
\label{TRASFEQ}
\end{array}
\end{equation}
where $\overline{  \log \frac{\rm  R^R_e}{ \rm R^K_e}  }$ is  the mean
logarithmic  ratio of  R-  and K-band  effective  radii, $\Gamma$  and
$\Delta$  are the  zeropoint  and the  slope~\footnote{  We note  that
Eq.~\ref{TRASFEQ}  was derived  assuming that  eq.~1 still  holds once
that aperture colours  are replaced by the total  galaxy colours, $\rm
R_T-K_T$.   We verified,  in fact,  that  the values  of $\Gamma$  and
$\Delta$  do not  change  significantly by  using  Kron magnitudes  to
estimate  the galaxy  colours.  }   of the  $\rm R-K$  vs.   $\rm K_T$
colour  magnitude  relation,  $\rm  \alpha'_K  =  \left(  1  +  \Delta
\right)^{-1} \cdot \left[  \alpha_R - \Gamma + 2.5  \cdot \Delta \cdot
\log  2  \pi  - \left(  5  -  \beta_R  \right) \cdot  \overline{  \log
(R^R_e/R^K_e)  }   \right]$,  $\rm  \beta'_K  =  \left(   1  +  \Delta
\right)^{-1}  \cdot  \left( \beta_R  +  5  \Delta  \right)$, and  $\rm
\gamma'_K  = -  \left( 5  - \beta_R  \right) \cdot  \left( 1  + \Delta
\right)^{-1}$.  We note that  the value of $\rm \alpha'_K$ corresponds
to the zeropoint of the NIR  KR, while $\rm \beta'_K$ is equal to $\rm
\beta_K$   only  if   the   third   term  of   the   last  member   of
Eq.~\ref{TRASFEQ} vanishes,  that is  if the ratio  of optical  to NIR
effective  radii  and,  therefore,  the internal  colour  gradient  of
galaxies  does  not  vary   systematically  along  the  \lre--\mie  \,
sequence.  By  using the  coefficients of the  CM relation  derived in
Sec.3.1,  the mean  ratio of  optical  to NIR  effective radii  ($\sim
0.16$),   and   the  R-band   KR   coefficients  ($\alpha=19.16$   and
$\beta=2.92$),  we obtain  $\rm \alpha'_K=15.57$,  $\rm \beta'_K=2.76$
and $\rm  \gamma'_K=-2.23$.  The value  of $\rm \alpha'_K$ is  in good
agreement  with  that of  $\rm  \alpha_K$,  while  the value  of  $\rm
\beta'_K$  is only marginally  consistent with  $\rm \beta_K$,  with a
difference  of $2.3\sigma$.   Since the  value of  $\rm  \gamma'_K$ is
negative, this difference could  be explained if the $\rm R_e^R/R_e^K$
ratio would become smaller for  larger values of $\rm R_e^K$, implying
that   larger   galaxies   have   a  less   steep   colour   gradient.
Fig.~\ref{CORPAR} plots $ \rm \log (R^R_e/R^K_e) $ and $\rm grad(R-K)$
as a function of $\rm \log R^K_e$.  Although the figure seems to favor
the trend suggested by the difference of $\beta'_K$ and $\beta_K$, the
error bars  do not allow  such a trend  to be definitively  proven. We
point  out,   however,  that  the   previous  results  are   in  clear
disagreement  with  a possible  steepening  of  colour gradients  with
galaxy size, since it would make the discrepancy between $\rm \beta_K$
and $\rm  \beta'_K$ strongly significant.  This point  will be further
addressed in Sec.~\ref{CONC}.


\begin{figure*}
  \centering
  \includegraphics[angle=0,width=11cm,height=11cm]{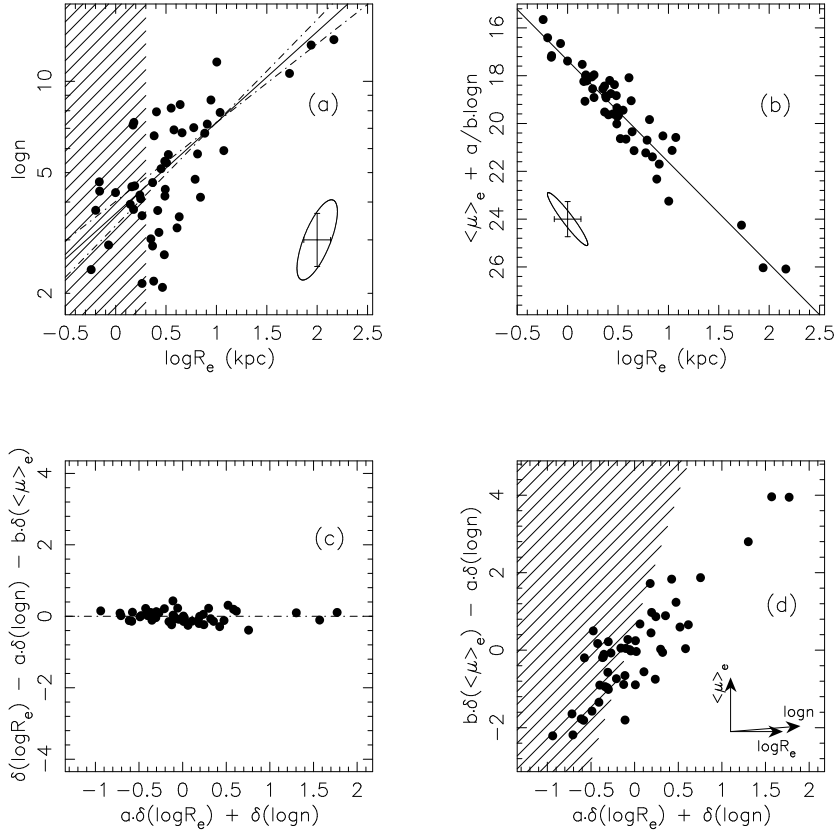}
  \caption{  Distribution  of  galaxies  in the  space  of  structural
   parameters.  Upper--left: \lRe--\lns  \, relation.  The ellipses is
   in the upper plots are  the $68\%$ probability contours.  The solid
   and  dot--dashed   lines  are  the  best--fit   and  the  $1\sigma$
   confidence  lines,  respectively.    The  region  affected  by  the
   magnitude selection  is filled by a  dashed pattern.  Upper--right:
   edge-on projection of the PHP  with the best-fit relation marked by
   a solid  line.  Lower--left: 'long' edge-on projection  of the PHP.
   Lower--right:  face--on view  of  the PHP.   The  shaded region  is
   delimited  on  the  right  by   the  line  that  results  from  the
   intersection of the plane $\rm K_T=const.$ with a plane parallel to
   the PHP but with zeropoint $\rm c+2 \cdot \sigma_c$.}
  \label{PPFIG}
\end{figure*}

\section{The NIR Photometric Plane at $\rm z \sim 0.2$}
\label{PPSEC}
Fig.~\ref{PPFIG}  plots different projections  of the  distribution of
galaxies in  the space  of the NIR  structural parameters\footnote{The
waveband dependence of the Photometric Plane will be investigated in a
forthcoming paper, by  using a larger sample of  galaxies.}.  As shown
in  Fig~\ref{PPFIG}a,  effective  radii  are  correlated  with  Sersic
indices  (e.g.  \citealt{YoC01}), in  the sense  that galaxies  with a
larger size tend  to have a surface brightness  profile more peaked in
the center.  By looking at Fig.~\ref{KRPLOT}, we see that, taking into
account  the dispersion  of the  KR, the  completeness  magnitude line
begins to intercept  the \lre--\mie \, relation at  $\rm \log R_e \sim
0.3$, implying  that galaxies in  the shaded area  of Fig~\ref{PPFIG}a
start  to be lost  from the  present sample  because of  the magnitude
selection.  We applied  a bi-weight least square fit  to the data with
$\rm \log R_e > 0.3$, minimizing the rms of the residuals with respect
to \lns.  The  fit gives $\rm \log n = (0.31\pm0.05)  \cdot \log R_e +
(0.56 \pm 0.1)$, with an rms that amounts to $\rm \sim 0.15dex$ ($\sim
35\%$) in \ns  \, and $\rm \sim 0.5dex$ in \lre,  which is much larger
with respect to that obtained for the KR ($\rm \sim 0.22\pm0.04~dex$).
We note that the \lns  \, regression is quite insensitive to selection
cuts with respect to \lre, and, in fact, the above coefficients do not
vary  significantly including  all the  galaxies in  the fit.   \\ The
other  panels of  Fig.~\ref{PPFIG} show  different projections  of the
so-called  Photometric Plane  (PHP).   For comparison  with GRA02,  we
adopted the following representation of the PHP:
\begin{equation}
\mathrm{
 \log R_e = a \cdot \log n + b < \! \mu \! >_e + c,
}  
\label{PPEQ}
\end{equation} 
which  is analogous  to the  usual equation  of the  Fundamental Plane
(FP),  once the velocity  dispersion term  is replaced  with $\mathrm{
\log  n  }$.   Following  GRA02,  the  coefficients  $\mathrm{a}$  and
$\mathrm{b}$ were obtained by a least square fit minimizing the rms of
the  residuals to  $\rm  \log R_e$.   We  also took  into account  the
correlation of the uncertainties on the structural parameters by using
the  MIST  algorithm  \citep{LBC00}.   The fitting  results  are  $\rm
a=0.6\pm0.13$,  $\rm b=0.235\pm0.02$ and  $\rm c=-4.5\pm0.25$,  with a
residual rms of $\rm \sim0.15~dex$  in $\rm R_e$.  In order to correct
these values for the magnitude cut,  we generalized to the 3D case the
procedure adopted in  the previous section for the  KR fit (see LBM03b
for details).   We found that the  bias in the $\rm  b$ coefficient is
negligible,  while   it  amounts  to   $+30\%$  and  $-4\%$   for  the
coefficients $\rm  a$ and $\rm c$,  respectively, and to  $+13 \%$ for
the dispersion in the $\rm  \log R_e$ variable.  After correction, the
fitting  coefficients become  $\rm a=0.8\pm0.2$,  $\rm b=0.235\pm0.02$
and $\rm c = -4.70\pm0.026$, with  a corrected \lre \, scatter of $\rm
\sim 0.17~dex$,  which is smaller by  $\sim14\%$ than that  of the KR.
The `corrected' values  of a, b and c were used  to obtain the edge-on
sections   (Fig.~\ref{PPFIG}b,   c)   and   the   face-on   projection
(Fig.~\ref{PPFIG}d) of the  PHP. \\ The slopes and  the scatter of the
NIR PHP  at $\rm z  \sim 0.2$ turn  out to be in  remarkable agreement
with  those derived  by GRA02  for galaxies  in the  Fornax  and Virgo
clusters: $\rm  a = 0.86 \pm  0.13$, $\rm b =  0.228\pm0.036$ and $\rm
\sigma_{\log  R_e} \sim 0.17~dex$.   We point  out that  the magnitude
selection  significantly affects  the  \lns \,  coefficient and  that,
therefore, it  would be very interesting  to use a  sample of galaxies
spanning a  larger magnitude range  in order to better  constrain this
coefficient.   The selection  effects  can be  further illustrated  by
looking  at the face-on  projection of  the PHP  in Fig.~\ref{PPFIG}d.
Due  to the  finite dispersion  of the  galaxies around  the  PHP, its
intersection with the plane $\rm K_T=cost$ results in a `strip', whose
`upper'   border   marks   the   limits   of  the   dashed   area   in
Fig.~\ref{PPFIG}d.   As consequence, the  distribution of  galaxies in
this area  is not 'complete'.  \\  In order to  estimate the intrinsic
dispersions  of  the  \lRe--\lns  \,  relation  and  of  the  PHP,  we
subtracted to the observed dispersion in \lRe \, the amount of scatter
expected  from  the measurement  errors  on  the observed  parameters,
taking into account the  covariance terms between the uncertainties on
\lRe, \mie \,  and \lns.  The intrinsic dispersions  of the \lRe--\lns
\, relation and of the PHP  turn out to be $\rm \sim0.45~dex$ and $\rm
\sim0.16~dex$,   respectively.   Interestingly,   we  find   that  the
measurement  errors account  only  for few  percents  of the  observed
dispersion  around the  PHP, because,  as shown  in Fig.~\ref{PPFIG}b,
their correlation is almost parallel to  the plane, in the same way as
for the KR.

\section{Discussion and conclusions.}
\label{CONC}
We have studied the galaxy population in a $5' \times 5'$ field around
the X-ray extended  source \A2163B, at about $6.5'$  North the cluster
of galaxies \A2163 ($\rm z=0.201$).  The BVRIK catalogue is presented,
including total magnitudes, colours, photometric redshifts and the RIK
structural parameters  (effective radius, mean  surface brightness and
Sersic index).

The  K-band luminosity  density  map  of \A2163B  \,  presents a  main
central  overdensity  of radius  $\rm  \sim0.2~Mpc$,  and a  secondary
structure  $\rm \sim0.4~Mpc$  from the  center.  The  colour magnitude
diagrams  show a  sharp red  sequence whose  slope flattens  at longer
wavelengths:  galaxy colours  become  bluer by  $\rm \sim0.1~mag$  per
magnitude  in  the $\rm  B-K$  vs. $\rm  K_T$  diagrams,  and by  $\rm
\sim0.04~mag$ into the  $\rm I-K$ vs.  $\rm K_T$  plane.  These values
agree  with the  predictions of  a  mass-metallicity model  of the  CM
relation (KAB98).  By introducing  a suitable procedure to account for
the  Milky  Way reddening,  we  find that  the  zeropoints  of the  CM
sequence  are  consistent  with   those  expected  for  a  red  galaxy
population at redshift $\rm z_c  \sim 0.2$, implying that \A2163B is a
cluster  of   galaxies  likely  involved  in  a   merging  event  with
\A2163. The  value of $\rm z_c$  is also confirmed by  the analysis of
the K-band LF and by the photometric redshift technique, which is used
to select  cluster members.  The luminosity segregation  of \A2163B \,
is investigated  by means  of the  K-band LF.  The  global LF  is well
described  by a single  Schecter function  with a  faint end  slope of
$\alpha= -1.2$, in agreement with other  studies of the NIR LF both in
clusters  and  in  the   field  (e.g.   \citealt{AND01},  Kochanek  et
al.~2001).   On  the  other  hand,  the faint  end  slope  depends  on
environment, varying from about $-1$ in the two higher density clumps,
to about $-1.4$ into the outer region, implying that the population of
dwarf  galaxies is less  abundant into  the denser  environment.  This
behaviour  of the  LF  suggests  that some  mechanism,  such as  tidal
disruption  or   cannibalism,  is   working  in  the   higher  density
environments,  decreasing  the number  of  faint  galaxies (Haines  et
al.~2003, submitted).

Surface photometry has been derived in the R, I and K bands for $\rm N
\sim 60$ cluster galaxies, and has been used:
\begin{description}
\item[i)] to compare galaxy parameters at optical/NIR wavelengths, and
 to  estimate  the  optical  and  NIR  internal  colour  gradients;
\item[ii)] to perform a comparative optical/NIR analysis of the KR;
\item[iii)] to derive the PHP of  spheroids at $\rm z \sim 0.2$ in the
K band.
\end{description}
The characteristics of the light distribution in galaxies are the same
between the optical wavebands, while a significant difference is found
when comparing  optical to NIR  parameters.  Effective radii  become a
factor $1.45 \pm 0.15$ greater from the K to the R band, implying that
the light  profile in  galaxies is much  more concentrated in  the NIR
than in the  optical.  On the other hand,  surface brightnesses show a
waveband dependence that  is fully explained by the  behaviour of $\rm
R_e$  and  by  the  $\rm  R-K$  galaxy colours.   The  value  of  $\rm
R_e^{OPT}/R_e^{NIR}$  is consistent  with  what we  found in  previous
works on the optical/NIR  structural parameters of cluster galaxies at
$\rm  z\sim0.3$ and  $\sim z\sim0.6$  (see LBM02  and LBM03a),  and is
within the range of results reported in previous works.  \citet{NSZ02}
and \citet{RPD02} found $\rm  R_e^{OPT}/R_e^{NIR} \sim 2$ by analysing
brightest  cluster  galaxies at  $0.4<z<0.8$  and nearby  early-types,
respectively, while a  lower value, $\rm R_e^{OPT}/R_e^{NIR} \sim1.2$,
was found by \citet{PCD98}.  \\ The population of spheroids in \A2163B
\,  has  null  colour  gradients  at optical  wavelengths  ($\rm  V-R$
restframe) and  optical-NIR gradients that  are significantly negative
($\sim  \rm  -0.48\pm0.06~mag/dex$),   consistently  with  the  quoted
differences  among  the effective  radii.   As  shown  by LBM03a,  the
optical-NIR  gradients  do  not  evolve significantly  with  redshift,
amounting to  $-0.4 \pm 0.1$. As  detailed in that  work, the redshift
dependence  of  both   optical-NIR  and  UV-optical  colour  gradients
strongly constrains  the age  and metallicity of  the inner  and outer
galaxy stellar populations.  Using the colour gradient model of LBM03a
(see  model   Z1,  Tab.~6),  the  quoted  colour   gradients  imply  a
metallicity gradient in spheroids of  $\rm -0.2 \pm 0.1dex$ per decade
of radius. This  result is in agreement with  that reported from other
studies of colour gradients  for both nearby and intermediate redshift
galaxies (\citealt{SMG00, IMP02}).

Monolithic collapse  models predict that early-type  galaxies are more
metal  rich  in the  center  than  in  the outskirts.   During  galaxy
formation, the  gas dissipates its  kinetic energy carrying  the heavy
elements   ejected   from   the   evolved   stars   into   the   inner
region~\citep{LAR74}.  Numerical simulations predict a metal abundance
gradient of about  $-0.5$ (e.g.  \citealt{CAR84}, and \citealt{KAW01},
see their models B1/B2 which  correspond to the magnitude range of our
samples).   This value is  not consistent  with our  results, although
some caution  is needed  due to scatter  present both in  the observed
data  and  in  the  simulation  results  (see  \citealt{KAW01}).   The
difference  of observed and  predicted metallicity  gradients suggests
that gas dissipation  has a minor role in  monolithic collapse models.
We note, however,  that dissipation is required in  order to reproduce
the  mass-metallicity  relation   of  early-type  galaxies  and  their
moderate orbital anisotropies.  We  argue, therefore, that some tuning
of  the parameters  driving the  physics of  gas dissipation  would be
required  in  order  to  fit  all the  observed  properties  into  the
monolithic scenario.  A likely explanation of the observed metallicity
gradients resides in  the merging processes, which mix  the SPs inside
galaxies.   It has  been shown  that both  dissipationless  merging of
spheroidal  systems  \citep{WHI80}  and  dissipative merging  of  disk
galaxies  \citep{BeS01}   produce  more  shallow   stellar  population
gradients.

The slope and the zeropoint  of the optical \lre--\mie \, relation are
consistent with those  obtained in LBM03b.  For what  concerns the NIR
KR, we  compared its  slope and zeropoint  with the optical  values by
taking into  account (i)  the colour magnitude  relation and  (ii) the
mean value of the ratio of optical/NIR effective radii.  These factors
fully  explain the difference  among the  optical and  NIR zeropoints.
The  most interesting  result  is that  the  slope of  the R-band  KR,
corrected for  the effects (i)  and (ii), is lower  by $\sim2.3\sigma$
than the K-band value.  We  show that this difference can be explained
if the ratio of optical to NIR effective radii (the optical/NIR colour
gradient) {\it  decreases (flattens)}  for larger galaxies.   A direct
inspection of the $\rm \log R_e^R / R_e^K$ and the $\rm grad(V-K)$ vs.
$\rm \log R_e^K$  diagrams shows that such trend  exists.  Our data do
not favour  a steepening  of colour gradients  with galaxy  size, that
would imply a larger discrepancy  between the R- and K-band KR slopes.
A {\it  steepening} of colour gradients with  galaxy luminosity (size)
is a natural expectation of  the monolithic collapse model, due to the
galactic wind  mechanism \citep{LAR74}. The galactic  wind blows early
in less  massive galaxies, preventing  gas dissipation to  carry heavy
metals in the center, and  producing, therefore, a less steep gradient
in  these  systems.  A  steepening  of  colour  gradients with  galaxy
luminosity has  been unfruitfully looked for by  various studies (e.g.
\citealt{PDI90},  \citealt{BRG01}).  Recently,  \citet{TaO03} measured
the optical  colour gradients for $\rm  N=36$ galaxies at  $\rm z \sim
0$.   They found  that  the  optical colour  gradient  of very  bright
early-types ($\rm  L > L^\star$)  can steepen with  galaxy luminosity,
while an opposite trend was suggested for fainter galaxies.  A similar
bimodal behaviour was found  by \citep{BaP94} for the colour gradients
of early-type spiral bulges.  We note that our data do not allow us to
distinguish  the behaviour  of colour  gradients for  the  very bright
early-types, while they allow a larger magnitude range in both optical
and NIR to be analyzed.   In hierarchical merging models, more massive
galaxies form  from larger disk systems, which  have stronger internal
gradients.   Since merging  dilutes the  stellar  population gradients
\citep{WHI80}, if galaxies with  larger size have experienced a higher
merging rate \citep{CCD92}, their colour  profile can be as flat as or
even less steep than those of smaller systems.  The data of \A2163B \,
seem to favour this interpretation.

Finally,  we have analyzed  the so-called  Photometric Plane  (PHP) of
spheroids  at  $\rm z  \sim  0.2$.  We  find  that  galaxies follow  a
bivariate  relation  between   K-band  effective  parameters  and  the
logarithm of  the Sersic index: $\rm  \log R_e^K \propto  a \cdot \log
n^K + b  \cdot < \!  \mu \!  >_e^K$.   Accounting for the correlations
among the  uncertainties on structural  parameters as well as  for the
selection effects,  we find $\rm a =  0.8 \pm 0.2$ and  $\rm 0.235 \pm
0.02$, and an  observed dispersion of $\rm \sim0.17~dex$  in $\rm \log
R_e$.  These  values are consistent with the  optical PHP coefficients
found by  GRA02 for early-types in  the Fornax and  Coma clusters.  We
have  discussed how  the  magnitude cut  affects  the distribution  of
galaxies in the space of  the structural parameters: the $\log n$ term
of the  PHP turns  out to be  significantly affected by  the magnitude
selection, indicating that  it would be very interesting  to analyze a
larger,   deeper  sample   of  galaxies   to  further   constrain  the
coefficients  of  this  relation.   Interestingly, we  find  that  the
correlation  of  the uncertainties  on  the  structural parameters  is
almost parallel  to the PHP, analogously  to what happens  for the KR.
This implies  that the intrinsic scatter  of the PHP  amounts to about
$\rm 0.16~dex$, which  is only few percents smaller  than the observed
scatter,  but  $\sim13\%$  lower  than  that  of  the  KR  ($\rm  \sim
0.21~dex$).  Therefore, although the observed scatter around the plane
seems to  be similar  to that of  the spectroscopic  Fundamental Plane
($\rm \sim0.15~dex$) as found by GRA02, the present data indicate that
its intrinsic dispersion can be significantly larger.  This point will
have to be further analyzed  by studying samples of galaxies for which
both velocity dispersions and Sersic indices are available.

\appendix

\section{Data reduction}

\subsection{BVRI data}
\label{BVRIDATA}
Both the  images of the  first and of  the second run were  reduced by
following the same procedure.  After bias subtraction, the images were
divided  by a  flat-field  frame obtained  by  combining twilight  sky
exposures.  Since  this procedure did  not fully remove  low frequency
variations  of the  background across  the images  (at level  of $\sim
5\%$), we  obtained a super--flat  correction for each band  by median
combining the corresponding scientific exposures. A polynomial fit was
applied to each super--flat frame and the fitted surfaces were used to
correct the  scientific exposures.  This  procedure reduced background
variations in the  final images at the level of  $\sim 0.5 \%$.  After
flat--field, the cosmic  rays in each exposure were  detected by using
the IRAF task  COSMICRAYS, and a mask frame  was constructed including
hot  pixels and  the bad  columns. The  images of  each run  were then
combined  by using the  IRAF task  IMCOMBINE.  Since  only Run  II was
photometric, the  images of  Run I were  suitably scaled  and combined
with those of Run II. The  scaling factors were computed from the mean
difference  of  the magnitudes  of  bright  unsaturated  stars in  the
observed  field. The  seeing  FWHM is  $1.5''$,  $1.2''$, $1.0''$  and
$1.0''$ for the  B, V, R, and I bands,  respectively.  The images were
calibrated into  the Johnson-Kron-Cousins photometric  system by using
standard  stars  from  \citet{LAN93}.   Since  comparison  fields  for
different  airmasses were  not  available, we  adopted the  extinction
coefficients typical  for the site of La  Silla: $\mathrm{A_B=0.21, \,
A_V=0.13,   \,  A_R=0.09,  \,   A_I=0.05~mag/airmass}$.   Instrumental
magnitudes for the  Landolt stars were computed within  an aperture of
diameter $10''$ by means of S-Extractor \citep{BeA96}.  For each band,
in  order  to  obtain  the  zeropoint  and  the  colour  term  of  the
photometric   calibration,  we   performed  a   robust   least  square
fit. Colour terms were found to be small and were neglected, while the
zeropoints turned out to  be $\mathrm{ZP_B=24.31 \pm 0.01}$, $\mathrm{
ZP_V=24.92  \pm  0.01}$, $\mathrm{  ZP_R=25.06  \pm 0.01}$,  $\mathrm{
ZP_I=24.41 \pm 0.01~mag}$ (scaled to $\rm 1s$ exposure time).

\subsection{K-band data}
\label{KDATA}
For each  dithering sequence, the  exposures were dark  subtracted and
corrected for flat-field, using a  super flat frame obtained by median
combining all the images taken during the night. After this procedure,
the magnitudes of the standard stars showed a rms variation across the
chip  of  $\rm \sim  0.04~mag$.   To  achieve  a better  accuracy,  we
obtained an  illumination correction frame by  measuring the magnitude
of a  standard star at different positions  in a $5 \times  5$ grid on
the frame.  All the images were divided by the illumination correction
frame, which allowed the low  frequency component of the flat-field to
be corrected at  better than $1 \%$.  Since sky  subtraction is a very
troublesome step for  the reduction of NIR data,  particularly in high
density regions  mostly populated  by bright early-type  galaxies with
extended  halos, this point  was carefully  dealt with  by a  two step
procedure.  First,  each exposure was sky subtracted  by computing the
sky  frame  from  the median  of  the  six  closest frames  along  the
sequence.   The images were  then registered  with integer  shifts and
combined,   with   a  sigma   clipping   algorithm   for  cosmic   ray
rejection. This procedure, however,  overestimate the sky level in the
extended  halos of  galaxies, and  therefore,  it is  not suitable  to
derive the surface photometry.  For this reason, we reiterated the sky
subtraction, proceeding  as follows.  The first step  images were used
to  obtain  mask frames  for  the sources  in  the  field, by  running
S-Extractor with  the checkimage  OBJECTS option.  For  each sequence,
the mask  was expanded in  order to 'cover'  the galaxy halos  and was
de-registered to each dithered exposure, by including also hot and bad
pixels.   The sky  frames  were estimated  by  an average  of the  six
closest  frame for  each sequence  exposure, rejecting  masked pixels.
The   images   were   then    sky   subtracted   and   combined   with
IMCOMBINE\footnote{The  images taken under  non-photometric conditions
were suitably  scaled to  the others. }   (using the  SIGCLIP option),
resulting in  a final  image with a  seeing FWHM of  $\sim0.9''$.  The
photometric  calibration was  performed  into the  Ks standard  filter
\citep{PMK98},  deriving the instrumental  magnitudes of  the standard
stars within an aperture of diameter $8''$. The airmass correction was
performed   by  using   an  extinction   coefficient   $\mathrm{A_K  =
0.06~mag/airmass}$, which  was derived by comparing  the magnitudes of
bright  objects in  the  field for  different  airmasses.  The  K-band
zeropoint turned  out to  be $\mathrm{ ZP_K  = 22.422  \pm 0.015~mag}$
(scaled to $\rm 1~s$ exposure time).

\section{Deriving $\rm z_c$ from the CM relations}
\label{zcebv}
The zeropoints $\rm \Gamma_{K_r}$ of the CM relations (Eq.~\ref{CMEQ})
were fitted with the colours calculated for an old, passively evolving
SP, as a function of the  redshift $\rm z_c$ and of the reddening $\rm
E(B-V)$.  To account for the metallicity change along the CM sequence,
we proceeded as  follows.  As shown by \citet{MLM03},  the CM relation
is reproduced by the metallicity--luminosity relation:
\begin{equation}
\mathrm{
\log( Z / Z_\odot) = (-0.097 \pm 0.005) \cdot M_V + (-2.09 \pm 0.09).
}
\nonumber
\end{equation}
where $\rm M_V$ is the  absolute V-band magnitude.  Using the slope of
the CM relation  for nearby early-types from Bower  et al.~(1992), the
above relation  allows the absolute  K-band magnitude to  be estimated
for a given metallicity $\rm  Z$.  For $\rm Z=Z_\odot$, we obtain $\rm
K(Z_\odot)=K_\odot=-24.74\pm0.15$.  In  order to predict  the value of
$\rm K(Z_\odot)$ as a function of redshift ($\rm K_\odot(z)$), we have
to  take  into account  the  luminosity  distance  term, and  the  K+E
corrections.  To  obtain these  corrections, we used  galaxy templates
from  the  GISSEL00  code  (\citealt{BrC93}),  with a  Scalo  IMF,  an
exponential   star  formation  rate   $\rm  exp(-t/\tau)$   and  solar
metallicity.   These  models  were   also  used  to  predict  (1)  the
optical--NIR galaxy colours  as a function of the  redshift $\rm z_c$,
and  (2) the  absorption coefficients  $\rm A_N$  and $\rm  A_K$  as a
function of $\rm z_c$ and  $\rm E(B-V)$, using the extinction law from
SFD98.   The values  of $\rm  z_c$ and  $\rm E(B-V)$  were  derived by
minimizing the following expression:
\begin{equation}
\mathrm{ \chi^2 [ z_c, z_f, E(B-V) ] \! = \! \sum_M \left[ \Gamma_{K_\odot(z)}-
(N-K)_t + (A_N-A_K) \right]^2 }
\end{equation}
where  $\rm  (N-K)_t$ are  the  template  colours,  and $z_f$  is  the
formation  redshift, which  was  allowed  to vary  in  the range  $\rm
[1.2,10]$.   The uncertainties  on  $\rm z_c$  and  $\rm E(B-V)$  were
obtained by shifting the values of $\rm \Gamma_{K_\odot(z)}$ according
to   the  relative   uncertainties  and   re-iterating   the  $\chi^2$
minimization.  We  note that  the uncertainties on  $\rm \Gamma_{K_r}$
also include the contribution of the errors in photometric zeropoints,
which  were  added  in  quadrature  to  the  uncertainties  on  galaxy
magnitudes  (see Sec.~\ref{APSEC}).   We found  that the  estimates of
$\rm z_c$  and $\rm  E(B-V)$ do not  change significantly  by adopting
either  $\rm \tau=0.01~Gyr$,  (Simple Stellar  Population  models), or
$\rm \tau=1~Gyr$ (models with a more protracted star formation rate).


\begin{thebibliography}{}

  \bibitem[Andreon(2001)]{AND01} Andreon, S. 2001, ApJ, 547, 623

  \bibitem[Andreon \&  Pell\` o(2000)]{AnP00} Andreon,  S., \& Pell\'
    o, R. 2000, A\&A, 353, 479

  \bibitem[Arnaud et  al.(1994)]{AEB94} Arnaud,  M., Elbaz, D.,  B\" o
    hringer, H.,  Soucail, H., \& Mathez,  G. 1994, in  New Horizon of
    X-Ray Astronomy,  ed. F.  Makino  \& T.  Ohashi  (Tokyo: Universal
    Academy), 537

  \bibitem[Balcells   and  Peletier(1994)]{BaP94}  Balcells,   M.,  \&
  Peletier, R. F. 1994, AJ, 107, 135

  \bibitem[Barger      et     al.(1998)]{BAS98}      Barger,     A.J.,
    Arag\'on-Salamanca,  A.,  Smail,  I.,  Ellis, R.S.,  Couch,  W.J.,
    Dressler,  A.,  Oemler, A.,  Poggianti,  B.M.,  \& Sharples,  R.M.
    1998, \apj, 501, 522

  \bibitem[Bartholomew et  al.~(2001)]{BRG01} Bartholomew, L.J., Rose,
    J.A., Gaba, A.E., \& Caldwell, N.  2001, AJ, 122, 2913

  \bibitem[Beers, Flynn and Gebhardt(1990)]{BFG90} Beers, T.C.,
    Flynn, K., \& Gebhardt, K.  1990, AJ, 100, 32

  \bibitem[Bekki  \&   Shioya(2001)]{BeS01}  Bekki,  K.,   \&  Shioya,
  Y. 2001, Ap\&SS, 276, 767

  \bibitem[Bertin \& Arnout(1996)]{BeA96} Bertin, E., \& Arnout,
      S. 1996, A\&AS, 117, 393

  \bibitem[Bower, Lucey  and Ellis(1992)]{BLE92} Bower,  R.G., Lucey,
    J.R., \& Ellis, R.S. 1992, MNRAS, 254, 589

  \bibitem[Bruzual \& Charlot(1993)]{BrC93} Bruzual, G.A., \&
    Charlot, S., 1993, ApJ, 405, 538

  \bibitem[Burstein \&  Heiles(1984)]{BuH84} Burstein, D.,  \& Heiles,
    C. 1984, ApJS, 54, 33 (BH84)

  \bibitem[Busarello et al.(2002)]{BML02} Busarello, G., Merluzzi, P.,
    La Barbera, F., Massarotti, M., \& Capaccioli, M. 2002, A\&A, 389,
    787


  \bibitem[Caon,  Capaccioli  and  D'Onofrio(1993)]{CCD93} Caon,  N.,
    Capaccioli, M., \& D'Onofrio, M., 1993, MNRAS, 265, 1013

  \bibitem[Capaccioli,  Caon  and D'Onofrio(1992)]{CCD92}  Capaccioli,
    M., Caon, N., \& D'Onofrio, M. 1992, MNRAS, 259, 323

  \bibitem[Cardelli,  Clayton and  Mathis(1989)]{CAR}  Cardelli, J.A.,
    Clayton, G.C., \& Mathis, J.S. 1989, ApJ, 345, 245

  \bibitem[Carlberg(1984)]{CAR84} Carlberg, R.G. 1984, ApJ, 286, 403

  \bibitem[Charlot,  Worthey  and  Bressan(1996)]{CWB96} Charlot,  S.,
  Worthey, G., \& Bressan, A. 1996, ApJ, 457, 625

  \bibitem[Ciotti  and  Bertin(1999)]{CB99}  Ciotti,  L.,  \&  Bertin,
    G. 1999, A\&A, 352, 447



  \bibitem[de  Propris \&  Pritchet(1998)]{dPP98} de  Propris,  R., \&
    Pritchet, C.J. 1998, AJ, 116, 1118 (dPP98)

  \bibitem[de Propris  et al.(1999)]{dPS99} de  Propris, R., Stanford,
    S.A.,  Eisenhardt, P.R., Dickinson,  M., \&  Elston, R.  1999, AJ,
    118, 719 (dPS99)

  \bibitem[Elbaz,  Arnaud and  B\"  ohringer(1995)]{EAB95} Elbaz,  D.,
    Arnaud, M., \& B\" ohringer, H. 1995, A\&A, 293, 337 (EAB95)

  \bibitem[Feretti  et al.(2001)]{FFG01}  Feretti,  L., Fusco-Femiano,
    R., Giovannini, G., \& Govoni, F. 2001, A\&A, 373, 106 (FFG01)

  \bibitem[Garilli,  Maccagni and Andreon(1999)]{GMA99}  Garilli, B.,
    Maccagni, D., \& Andreon, S. 1999, A\&A, 342, 408

  \bibitem[Graham(2002)]{GRA02} Graham, A. 2002, 334, 859 (GRA02)

  \bibitem[Graham and Guzm\`an(2003)]{GrG03}  Graham, A., \& Guzm\`an,
    R. 2003, \aj, 125, 2936

  \bibitem[Haines  et al.(2003)]{HAI03}  Haines,  C.P., Mercurio,  A.,
  Merluzzi, P., et al. 2003, A\&A, submitted

  \bibitem[Huang et al.(2001)]{HTK01}  Huang, J.-S., Thompson, D., K\"
    ummel, M.W., et al. 2001, A\&A, 368, 787

  \bibitem[Idiart,  Michard and  Pacheco(2002)]{IMP02}  Idiart, T.P.,
    Michard, R., \& de Freitas Pacheco, J.A.  2002, A\&A, 383, 30

  \bibitem[J\o  rgensen, Franx  and  Kj\ae rgaard(1995)]{JFK95a}  J\o
    rgensen, I.,  Franx, M., \&  Kj\ae rgaard, P.  1995,  \mnras, 273,
    1097


  \bibitem[Kauffmann(1996)]{KAU96} Kauffmann, G. 1996, MNRAS, 281, 475

  \bibitem[Kawata(2001)]{KAW01} Kawata, D.  2001, ApJ, 558, 598

  \bibitem[Kochanek et al.(2001)]{KPF01}  Kochanek, C.S., Pahre, M.A.,
    Falco, E.E., et al.  2001, ApJ, 560, 566

  \bibitem[Kodama  et  al.(1998)]{KAB98}   Kodama,  T.,  Arimoto,  N.,
    Barger, A.J, \& Arag\'on-Salamanca, A. 1998, A\&A, 334, 99 (KAB98)


  \bibitem[Kron(1998)]{KRO80} Kron, R.G. 1980, ApJS, 43, 305

  \bibitem[La  Barbera,   Busarello  and  Capaccioli(2000)]{LBC00}  La
  Barbera, F., Busarello, G., \& Capaccioli, M. 2000, A\&A, 362, 851

  \bibitem[La Barbera et  al.(2002)]{LBM02} La Barbera, F., Busarello,
    G., Merluzzi,  P., Massarotti, M.,  \& Capaccioli, M.   2002, ApJ,
    571, 790 (LBM02)

  \bibitem[La   Barbera  et   al.(2003a)]{LBM03a}   La  Barbera,   F.,
    Busarello,  G.,   Massarotti,  M.,  Merluzzi,   P.,  \&  Mercurio,
    A. 2003a, \aap, 409, 21 (LBM03a)

  \bibitem[La   Barbera  et   al.(2003b)]{LBM03b}   La  Barbera,   F.,
    Busarello,  G.,  Massarotti,  M.,  Merluzzi,  P.,  \&  Capaccioli,
    M. 2003b, ApJ, 595, 127 (LBM03b)

  \bibitem[La Barbera et al.(2003c)]{LMI03c} La Barbera, F., Merluzzi,
    P., Iovino,  A., Busarello, G., Massarotti, M.,  \& Capaccioli, M.
    2003c, \aap, 399, 899

  \bibitem[Landolt(1992)]{LAN93} Landolt, A.U. 1992, AJ, 104, 340
 
  \bibitem[Larson(1974)]{LAR74} Larson, R.B., 1974 MNRAS, 166, 585


  \bibitem[Lubin and Sandage(2001)]{LuS01} Lubin, L.M., \& Sandage, A.
    2001, \apj, 122, 1084

  \bibitem[Markevitch et al.(1996)]{MMI96} Markevitch, M., Mushotzky,
    R., Inoue, H., et al. 1996, ApJ, 456, 437 



  \bibitem[Massarotti, Iovino and Buzzoni(2001)]{MAa} Massarotti, M.,
    Iovino, A., \& Buzzoni, A. 2001, A\&A, 368, 74

  \bibitem[Massarotti et  al.(2001)]{MAb} Massarotti, M.,  Iovino, A.,
    Buzzoni, A., \& Valls-Gabaud 2001, A\&A, 380, 425

  \bibitem[Mercurio et al.(2003)]{MMM03} Mercurio, A., Massarotti, M.,
  Merluzzi, P.,  Girardi, M., La  Barbera, F., \& Busarello,  G. 2003,
  A\&A, 408, 57
   
  \bibitem[Merluzzi et al.(2003)]{MLM03} Merluzzi, P., La Barbera, F.,
  Massarotti, M., Busarello, G., \& Capaccioli, M. 2003, ApJ, 589, 147

  \bibitem[Nelson et al.(2002)]{NSZ02} Nelson, A.E., Simard, L.,
   Zaritsky, D., Dalcanton, J.J., \& Gonzalez, A.H.  2002, ApJ, 567,
   144

  \bibitem[Pahre,  de  Carvalho  and  Djorgovski(1998)]{PCD98}  Pahre,
    A.M., de Carvalho, R.R., \& Djorgovski, S.G. 1998, AJ, 116, 1606

  \bibitem[Pahre,  Djorgovski  and  de Carvalho(1996)]{PDdC96}  Pahre,
    M.A., Djorgovski, S.G., \& de Carvalho, R.R. 1996, 456, 79


  \bibitem[Peletier et al.(1990)]{PDI90} Peletier, R.F., Davies, R.L.,
  Illingworth, G.D., Davis, L.E., \& Cawson, M.  1990, AJ, 100, 1091

  \bibitem[Peletier,  Valentijn  and  Jameson(1990)]{PVJ90}  Peletier,
    R.F., Valentijn, E.A., \& Jameson, R.F. 1990, A\&A, 233, 62

  \bibitem[Persson et al.(1998)]{PMK98} Persson, S.E., Murphy, D.C.,
   Krzeminski, W., Roth, M., \& Rieke, M.J. 1998, AJ, 116, 2475



  \bibitem[Pozzetti,  Bruzual and Zamorani(1996)]{POZ}  Pozzetti, L.,
    Bruzual A.G., \& Zamorani, G. 1996, MNRAS, 281, 953

  \bibitem[Press \& Schechter(1974)]{PrS74} Press, W.H., \& Schechter,
  P. 1974, ApJ, 187, 425

  \bibitem[Rembold  et   al.(2002)]{RPD02}Rembold,  S.B.,  Pastorizia,
   M.G., Dicati, J.R., Rubio, M., \& Roth, M. 2002, A\&A, 391, 531

  \bibitem[Saglia et al.(2000)]{SMG00} Saglia, R.P., Maraston, C.,
    Greggio, L., Bender, R., \& Ziegler, B. 2000, A\&A, 360, 911

  \bibitem[Sandage   and  Perelmuter(1991)]{SaP91}  Sandage,   A.,  \&
    Perelmuter, J.-M. 1991, \apj, 370, 455

  \bibitem[Sandage  and   Lubin(2001)]{SaL01}  Sandage,A.,  \&  Lubin,
    L.M. 2001, \apj, 121, 2271

  \bibitem[Schlegel, Finkbeiner  and Davis(1998)]{SFD98} Schlegel, D.,
    Finkbeiner, D.P., \& Davis, M. 1998, ApJ, 500, 525 (SFD98)


  \bibitem[Stanford, Eisenhardt and Dickinson(1998)]{SED98} Stanford,
    S.A., Eisenhardt, P.R.M., \& Dickinson, M. 1998, ApJ, 492, 461

  \bibitem[Tamura and Ohta(2000)]{TaO00} Tamura, N., \& Ohta, K. 2000,
    AJ, 120, 533

  \bibitem[Tamura and Ohta(2003)]{TaO03} Tamura, N., \& Ohta, K. 2003,
    AJ, 126, 596


  \bibitem[Visvanathan \& Sandage(1977)]{ViS77} Visvanathan, N., \&
    Sandage, A. 1977, ApJ, 216, 21

  \bibitem[Young  and  Currie(2001)]{YoC01}  Young, C.K.,  \&  Currie,
    M.J. 2001, A\&A, 369, 736

  \bibitem[White(1980)]{WHI80} White, S.D. 1980, MNRAS, 191, 1

  \bibitem[Worthey et al.(1996)]{WTF96}  Worthey, G., Trager, S.C., \&
    Faber,  S.M.,  1996,  in  ASP  Conf.  Ser.   86,  Fresh  Views  of
    Elliptical Galaxies, ed.  A.  Buzzoni, A.  Renzini, \& A.  Serrano
    (San Francisco: ASP), 203

  \bibitem[Ziegler et  al.(1999)]{ZSB99} Ziegler, B.L.,  Saglia, R.B.,
    Bender, R., et al. 1999, A\&A, 346, 13

\end{thebibliography}
\end{document}